\documentclass[apj]{emulateapj}

\usepackage{graphicx,times}
\usepackage{subfigure}
\newcommand{\be}{\begin{equation}}
\usepackage{threeparttable}
\usepackage{booktabs}
\newcommand{\ee}{\end{equation}}
\newcommand{\bea}{\begin{eqnarray}}
\newcommand{\eea}{\end{eqnarray}}

\usepackage{enumerate}
\usepackage{amsmath}
\usepackage{cases}
\usepackage{longtable}
\usepackage{hyperref}
\usepackage{epstopdf}
\usepackage{amsmath,bm}
\usepackage{amssymb}
\usepackage{natbib}
\usepackage{morefloats}
\usepackage{multirow}
\usepackage{array}
\usepackage{verbatim}

\begin{document}

\title{Turbulent dynamo in a weakly ionized medium}

\author{Siyao Xu\altaffilmark{1,2},
Sudip K. Garain\altaffilmark{3},
Dinshaw S. Balsara\altaffilmark{3},
and Alex Lazarian\altaffilmark{1} }

\altaffiltext{1}{Department of Astronomy, University of Wisconsin, 475 North Charter Street, Madison, WI 53706, USA; 
sxu93@wisc.edu,
lazarian@astro.wisc.edu}
\altaffiltext{2}{Hubble Fellow}
\altaffiltext{3}{Department of Physics, University of Notre Dame, Notre Dame, IN 46556, USA;
sgarain@nd.edu,
dbalsara@nd.edu}

\begin{abstract}

The small-scale turbulent dynamo is an important process contributing to the cosmic magnetization. 
In partially ionized astrophysical plasmas, 
the dynamo growth of magnetic energy strongly depends on the coupling state between ions and neutrals and 
the ion-neutral collisional damping effect. 
A new damping stage of turbulent dynamo in a weakly ionized medium was theoretically predicted by \citet{XL16}.
By carrying out a 3D two-fluid dynamo simulation, 
here we for the first time numerically confirmed the physical conditions and the 
linear-in-time growth of magnetic field strength of the damping stage of dynamo. 
The dynamo-amplified magnetic field has a characteristic length as the damping scale, 
which increases with time and can reach the injection scale of turbulence after around 
eight largest eddy-turnover times given sufficiently low ionization fraction and weak initial magnetic field. 
Due to the weak coupling between ions and neutrals, most turbulent energy carried by neutrals cannot be converted to the magnetic energy, 
resulting in a relatively weak magnetic field at the end of dynamo. 
This result has important implications for the growth of 
magnetic fields in the partially ionized interstellar medium and 
shock acceleration of Galactic cosmic rays.

\end{abstract}


\section{Introduction}

Magnetic fields pervade the Universe and are manifest in diverse astrophysical systems
\citep{Han17}.
The turbulent dynamo, which both amplifies the strength of the magnetic field and increases its coherence length,  
is the most promising mechanism to account for the growth and maintenance of the cosmic magnetism
\citep{Bran05}.
In particular, the turbulent dynamo acting on scales comparable or smaller than the driving scale of turbulence, i.e., the 
small-scale dynamo, is much more efficient than the large-scale dynamo, 
and also more generally operates in astrophysical environments 
wherever the turbulent energy exceeds the magnetic energy.

{Depending on the physical conditions, there are a variety of dynamo regimes 
(\citealt{XL16}, hereafter XL16).
In the case of a large Prandtl number, which is the ratio of viscosity to resistivity,} 
the kinematic regime of the small-scale dynamo at sub-viscous scales has been extensively studied 
(e.g., \citealt{Mar02,Sc02,Ma04}).
The concentration of the magnetic energy at the small resistive scale claimed in these theoretical and 
low-resolution numerical studies was disproved by high-resolution dynamo simulations 
\citep{Hau04}.
\footnote{In fact, after a close inspection of, e.g., 
figure 1 of \citet{Mar02} (as pointed out by \citet{Hau03}), 
figure 12 in \citet{Ma04}, 
one can easily see that their results also show the peak of the magnetic energy spectrum significantly away from the 
resistive scale. }
Meanwhile, the nonlinear regime of the small-scale dynamo in the inertial range of turbulence has also been studied numerically 
\citep{Cho00,CVB09,Bere11}, 
which is found to be characterized by a very inefficient linear-in-time growth of magnetic energy. 
Recent theoretical and numerical advances in the study of magnetohydrodynamic (MHD) turbulence 
\citep{GS95,LV99,MG01,CLV_incomp,KL09,KL12}
enable us to construct an analytical theory of the nonlinear turbulent dynamo 
(XL16), 
which has been shown in quantitative agreement with numerical measurements. 
In XL16, 
the turbulent diffusion of magnetic fields enabled by the turbulent magnetic reconnection 
\citep{LV99}
was identified as the physical origin of the low efficiency of the nonlinear dynamo. 
Besides, XL16 also analytically discovered
a transitional stage connecting the kinematic and nonlinear regimes, where the peak of the magnetic energy spectrum shifts from the 
resistive scale to the viscous scale. 
Their theoretical prediction on the 
sub-viscous spectral tail $k^{-1}$ formed during the transitional stage is consistent with the numerical result in 
\citet{Hau04}.

In astrophysical plasmas with a significant neutral component in e.g., the early Universe, 
cold phases of the interstellar medium (ISM), protoplanetary disks, the solar chromosphere, 
both MHD turbulence and turbulent dynamo are influenced by the partial ionization 
\citep{XLr17}.
Ion-neutral collisional damping of linear MHD waves has been earlier studied by, 
e.g., \citet{Lan78,Bals96,Zaqa11}.
On the basis of the updated understanding of MHD turbulence mentioned above, 
the damping of MHD turbulence due to ion-neutral collisions and the viscosity in neutrals 
has been studied both analytically 
\citep{LG01,LVC04,XLY14,Xuc16,XLr17}
and numerically
\citep{TB08,Till11,TilBal10,Mey14,Burk15}.

Regarding the small-scale dynamo in a partially ionized medium, the damping effect due to ion-neutral collisions on the efficiency of dynamo  
has been discussed in, e.g., 
\citet{KulA92, Sub98}.
The new findings in XL16 include
(i) a sub-viscous spectral tail $k^{-1}$ formed during the transitional stage at a relatively high ionization fraction; 
(ii) a damping stage of dynamo characterized by a linear-in-time growth of magnetic field strength at a relatively low ionization fraction; 
(iii) the nonlinear stage of dynamo with a universal dynamo efficiency irrespective of the ionization fraction;
(iv) a direct relation of the damping of MHD turbulence to that of turbulent dynamo. 
These theoretical findings have also been applied to studying the role of magnetic fields in, e.g. 
the star formation in the early Universe 
(XL16), 
cosmic ray acceleration at shocks
(\citealt{XL17}, hereafter XL17),

In this work, our purpose is to numerically test the damping stage of dynamo in a weakly ionized medium. 
Different from the exponential growth of magnetic energy in the sub-viscous range 
\citep{KulA92},
XL16 demonstrated that 
the damping stage of dynamo takes place within the inertial range of turbulence. 
It arises at a sufficiently low ionization fraction so that 
(a) ions and neutrals are only weakly coupled, and thus most turbulent energy in neutrals is not involved in the dynamo; 
(b) the ion-neutral collisional damping scale coincides with the dynamo driving scale; 
(c) the magnetic field strength grows linearly with time;
and (d) there is no equipartition between the turbulent and magnetic energies. 
We will present the first numerical test of the theoretical prediction on the damping stage of dynamo
in XL16 by carrying out a 3D two-fluid numerical simulation. 
We use the two-fluid version of the RIEMANN code
\citep{Bal98a,Bal98b,Bal04,Bal10,Bal12,BS99a,BS99b}
to simulate the weakly ionized turbulent plasma. 
The RIEMANN code has been widely used for studying astrophysical problems in partially ionized plasmas
(e.g., \citealt{TilBal10,Till11,Mey14}). 
In general, a two-fluid MHD simulation requires extensive computational effort. 
Our two-fluid dynamo simulation is even more challenging in order to achieve 
(i) a low ionization fraction to ensure the emergence of the damping stage; 
(ii) a large inertial range, as the damping scale increases with time; 
and (iii) a long simulation time to observe the entire dynamo evolution of magnetic fields. 
Despite its high computational cost, 
this numerical testing will provide direct evidence for the XL16 theory of the damping stage of dynamo 
and quantitatively reinforce our understanding of the dynamo physics in a weakly ionized medium. 
It is also important for further applications of the theory to studying the evolution and structure of magnetic fields in neutral dominated 
astrophysical environments.

The paper is organized as follows.
In Section 2, we describe the physical conditions and the analytically derived evolution law of the magnetic field 
for the damping stage of dynamo. 
In Section 3, we present the numerical results of the two-fluid simulation and their comparisons with our theoretical predictions.
In Section 4, we further examine the importance of the damping stage of dynamo in the partially ionized ISM. 
The discussion {about the effect of ion-neutral coupling on MHD turbulence and turbulent dynamo} is in Section 5. 
The summary follows in Section 6.

\section{Damping stage of dynamo in a weakly ionized medium}
\label{sec: damthe}

By stretching magnetic field lines, turbulent motions can amplify magnetic fields. 
Meanwhile, magnetic fields also undergo diffusion due to plasma or/and turbulence effects. 
These two opposing processes, turbulence stretching and magnetic field diffusion, together determine the dynamo efficiency.

In the kinematic dynamo regime, the magnetic energy is lower than the turbulent energy, and the magnetic field is 
dynamically unimportant. 
The diffusion only arises from plasma effects.
In the case of a weakly ionized plasma, {i.e., molecular clouds in the ISM,}
the diffusion in the kinematic dynamo regime mainly comes from the slippage between ions and neutrals.
{So the ion-neutral collisional damping is the dominant damping process of magnetic fluctuations, 
whereas other damping effects including the viscous damping and resistive damping are negligible
\citep{KulA92,Xuc16}.}

{Here we consider the damping stage of dynamo in a weakly ionized medium, 
which was first identified by 
XL16. 
It is in the kinematic regime and subjected to severe ion-neutral collisional damping. }

\subsection{Physical conditions for the damping stage of dynamo}
\label{ssec: con}

{Depending on the ionization fraction, the turbulent dynamo in a partially ionized medium undergoes different evolutionary stages. 
To observe a significant damping effect on the dynamo growth of magnetic energy in a damping stage, 
the ionization fraction should be sufficiently small, so that the 
ion-neutral coupling is weak and the ion-neutral collisional damping is strong. 
We note that unlike the strongly coupled regime where ions and neutrals are strongly coupled together
and the decoupled regime where the two species are decoupled from each other, 
in the weakly coupled regime considered here, neutrals are decoupled from ions, 
but ions can still collide with surrounding neutrals in a weakly ionized medium, and thus 
the motions of ions and magnetic fields are most severely damped 
\citep{Xuc16}.
Next we detail the physical conditions for the damping stage of dynamo to arise. }

\begin{longtable}{lc}
\caption[]{List of main notations
}\label{tab: nota} \\ 
\hline\hline
Description        & Symbol  \\
\hline
magnetic energy & $\mathcal{E}_M$ \\
magnetic energy spectrum & $M(k, t)$ \\
drag coefficient & $\gamma_d$ \\
neutral-ion collision frequency & $\nu_{ni}$ \\
ion-neutral collision frequency & $\nu_{in}$ \\
ion-neutral collisional damping scale & $l_d$ \\
neutral viscosity & $\nu_n$ \\
viscous damping scale & $l_\nu$ \\
peak scale of $M(k, t)$ & $l_p$ \\
injection scale of turbulence & $L$ \\
turbulent velocity at $L$ & $V_L$ \\
eddy-turnover time at $L$ & $\tau_\text{eddy}$ \\
turbulent velocity at $l$  & $v_l$ \\
stretching rate / turnover rate at length scale $l$ & $\Gamma_l$  \\
stretching rate / turnover rate at $l_d$ & $\Gamma_d$  \\
stretching rate / turnover rate at $l_\nu$ & $\Gamma_\nu$  \\
stretching rate / turnover rate at $L$ & $\Gamma_L$  \\
stretching rate / turnover rate at $l_p$ & $\Gamma_p$  \\
ion mass density & $\rho_i$ \\
neutral mass density & $\rho_n$ \\
total mass density & $\rho$ \\
neutral fraction & $\xi_n$ \\
ion-neutral coupling coefficient & $\eta_c$ \\
effective density (Eq. \eqref{eq: efdid}) & $\rho_\text{eff}$ \\
Alfv$\acute{e}$n speed of ionized fluid & $V_{Ai}$ \\
Alfv$\acute{e}$n Mach number of ionized fluid & $M_{Ai}$ \\
Alfv$\acute{e}$n speed of strongly coupled ions and neutrals & $V_{A,\text{tot}}$ \\
Alfv$\acute{e}$n speed in terms of $\rho_\text{eff}$ & $V_{A,\text{eff}}$ \\
\hline\hline
\end{longtable}

{\it Condition (1)}: a sufficiently small ionization fraction

The damping stage of dynamo is characterized by the 
weak coupling state between ions and neutrals and the consequent severe ion-neutral collisional damping. 
Quantitatively, the neutral-ion collisional frequency $\nu_{ni}$ should be smaller than the 
dynamo stretching rate $\Gamma_l$ of magnetic fields {to ensure the weak coupling between ions and neutrals 
(see Table \ref{tab: nota} for the main notations used in this paper)}.
The former is given by $\nu_{ni} = \gamma_d \rho_i$, with the drag coefficient $ \gamma_d$
(see e.g. \citealt{Shu92})
and the ion density $\rho_i$.  
The latter is determined by the 
turbulence eddy-turnover rate $v_l / l$, where $v_l$ is the turbulent velocity at the length scale $l$. 
According to the Kolmogorov scaling of hydrodynamic turbulence, $v_l$ decreases with $l$ as 
\begin{equation}\label{eq: kol}
    v_l = V_L \Big(\frac{l}{L}\Big)^\frac{1}{3}
\end{equation}
along the turbulent energy cascade, 
where $V_L$ is the turbulent velocity at the injection scale $L$ of turbulence. 
It can be easily seen that smaller eddies have larger eddy-turnover rates. 
Since the eddies at the {ion-neutral collisional} damping scale $l_d$ of magnetic fluctuations are the smallest ones that can effectively stretch 
magnetic field lines, they are mainly responsible for the dynamo action. 
The corresponding dynamo stretching rate is $\Gamma_d = v_d / l_d$, 
where $v_d$ is the turbulent velocity at $l_d$.

The above condition is formulated as 
(XL17)
\begin{equation}\label{eq: conlen}
   \frac{2 }{ \mathcal{C} \Gamma_d} < 1,
\end{equation}
where 
\begin{equation}\label{eq: parac}
    \mathcal{C} = \frac{\xi_n}{ 3 \nu_{ni}} \approx \frac{1}{3 \nu_{ni}},
\end{equation}
which imposes a constraint on the maximum value of the ionization fraction. 
We note that the expression on the LHS of Eq. \eqref{eq: conlen} is related to the Reynolds number at $l_d$ defined in 
\citet{Bals96}.
Here the ratio between the neutral density and the total density $\xi_n = \rho_n / \rho$
is approximately equal to unity in a weakly ionized medium. 
It implies that when the ionization fraction is sufficiently small, neutrals collide with ions so infrequently that 
neutrals are basically decoupled from the dynamo-stretched field lines. 
On the other hand, in a neutral dominated medium, ions can still collide with surrounding neutrals.  
Quantitatively, there is $\nu_{in} \gg \Gamma_d$, 
where $\nu_{in} = \gamma_d \rho_n$ is the ion-neutral collisional frequency. 
It is related to $\nu_{ni}$ by $\nu_{in} =(\rho_n/\rho_i) \nu_{ni} $. 
Evidently, $\nu_{in}$ is much larger than $\nu_{ni}$ in a weakly ionized medium. 
Because of the weak coupling between ions and neutrals, 
the dynamo action cannot effectively convert the 
turbulent kinetic energy carried by neutrals to the magnetic energy.

{\it Condition (2)}: sufficiently small magnetic energy

As mentioned earlier, the magnetic energy in the kinematic dynamo regime is smaller than the turbulent energy. 
At $l_d$, where the local turbulent motions dominate the dynamo action, there should be 
\begin{equation}\label{eq: con3}
    \mathcal{E}_M < \frac{1}{2} v_d^2 
\end{equation} 
where
\begin{equation}
     \mathcal{E}_M = \frac{1}{2} V_{A}^2 
\end{equation}
is the magnetic energy, and $V_A$ is the Alfv\'{e}n speed. 
So the relation in Eq. \eqref{eq: con3} is equivalent to $V_A < v_d$.

Meanwhile, there exists the equalization between $\Gamma_d$ and the ion-neutral collisional damping rate $\omega_\text{IN}$ at $l_d$,
where $\omega_\text{IN}$ is given by 
\citep{KulA92}
\begin{equation}\label{eq: omin}
     \omega_\text{IN} =  \mathcal{C} l^{-2} \mathcal{E}_M  = \frac{\mathcal{C}}{2}  l^{-2} V_A^2.
\end{equation}
From $\Gamma_d = \omega_\text{IN}$ at $l_d$, we find
\begin{equation}\label{eq: ldsip}
   l_d = \frac{\mathcal{C}}{2} V_A^2 v_d^{-1}. 
\end{equation}
Combining the above expression with the condition in Eq. \eqref{eq: conlen} yields $V_A < v_d$. 
It shows that under {\it Condition (1)}, {\it Condition (2)} is naturally satisfied.
In fact, due to the severe damping effect at a small ionization fraction, the equipartition between the magnetic and turbulent energies at $l_d$ cannot be 
reached. 
Any further growth of magnetic energy would break the balanced condition $\Gamma_d = \omega_\text{IN}$ at $l_d$
until the new balance is achieved at a larger $l_d$. 
Hence the dynamo in the damping stage remains in the kinematic regime.

{\it Condition (3)}: dominant ion-neutral collisional damping over the neutral viscous damping 

{As mentioned above, the ion-neutral collisional damping is the dominant damping effect for the damping stage of dynamo. 
But we note that as the ion-neutral collisional damping depends on the magnetic energy (Eq. \eqref{eq: omin}),
to ensure $\omega_\text{IN} > \omega_\text{NV}$,} 
where 
\begin{equation}\label{eq: sknu}
      \omega_\text{NV} = l^{-2} \nu_n, 
\end{equation}
is the damping rate related to the kinematic viscosity $\nu_n$ in neutrals, we should have the magnetic energy 
(Eqs. \eqref{eq: omin} and \eqref{eq: sknu})
\begin{equation}\label{eq: con2}
   \mathcal{E}_M > \mathcal{C}^{-1} \nu_n. 
\end{equation}
{When we consider a small ionization fraction and the dynamo growth of magnetic energy, 
the above condition can be easily satisfied.}

Alternatively, when the ion-neutral collisional damping dominates over the neutral viscous damping, $l_d$ should be 
larger than the viscous damping scale $l_\nu$. 
The condition $l_d > l_\nu$ yields (Eq. \eqref{eq: ldsip})
\begin{equation}
    \mathcal{C} \mathcal{E}_M v_d^{-1} > \nu_n v_\nu^{-1}, 
\end{equation}
where the relation $l_\nu^{-2} \nu_n = v_\nu / l_\nu$ is used, and $v_\nu$ is the turbulent velocity at $l_\nu$. 
Since $v_d > v_\nu$, there must be 
\begin{equation}
     \mathcal{E}_M  > \mathcal{C}^{-1} \nu_n, 
\end{equation}
which recovers the condition in Eq. \eqref{eq: con2}.

{Under the above conditions (Eq. \eqref{eq: conlen} and Eq. \eqref{eq: con2}), we expect that the turbulent dynamo in a weakly ionized medium 
undergoes a damping stage.}

\subsection{Magnetic field evolution during the damping stage of dynamo}
\label{ssec: the}

In the damping stage, the time evolution of magnetic fields strongly depends on the ion-neutral collisional damping. 
As mentioned earlier, the dynamo stretching rate is given by the eddy-turnover rate at $l_d$, 
\begin{equation}\label{eq: gamd}
   \Gamma_d = \frac{v_d}{l_d} = L^{-\frac{1}{3}} V_L l_d^{-\frac{2}{3}},
\end{equation}
where the Kolmogorov scaling in Eq. \eqref{eq: kol} is used. 
With the same scaling, the expression of $l_d$ in Eq. \eqref{eq: ldsip} can be rewritten as 
\begin{equation}\label{eq: lde}
   l_d = \mathcal{C}^\frac{3}{4} L^\frac{1}{4} V_L^{-\frac{3}{4}} \mathcal{E}_M^\frac{3}{4}. 
\end{equation}
The growth of $\mathcal{E}_M$ results in a stronger damping effect and a larger $l_d$.

The magnetic fluctuations on length scales larger than $l_d$ follow the Kazantsev spectrum
\citep{Kaza68}
as a result of the dynamo stretching, 
\begin{equation}\label{eq: mk}
   M(k, t) = M_1 \exp \bigg(\frac{3}{4} \int \Gamma_d dt \bigg) \bigg(\frac{k}{k_1}\bigg)^\frac{3}{2}, 
\end{equation}
where $M_1$ is the initial magnetic energy spectrum at some reference wavenumber $k_1$. 
The Kazantsev spectrum has dependence on both wavenumber $k$ and time $t$. 
By integrating $M(k, t)$ over $k$, we can derive $\mathcal{E}_M$ as a function of $t$,
\begin{equation}\label{eq: inmke}
      \mathcal{E}_M(t) = \frac{1}{2} \int_0^{k_d} M(k, t) dk. 
\end{equation}
Combining Eqs. \eqref{eq: gamd}-\eqref{eq: inmke} and after some straightforward algebra, we arrive at 
(XL16),
\begin{equation}\label{eq: emdt}
   \sqrt{ \mathcal{E}_M} = \sqrt{ \mathcal{E}_{M1}} + \frac{3}{23} \mathcal{C}^{-\frac{1}{2}} L^{-\frac{1}{2}} 
   V_L^\frac{3}{2} (t - t_1) ,
\end{equation}
with the magnetic energy $\mathcal{E}_{M1}$ at the beginning of the damping stage $t = t_1$. 
As $\sqrt{ \mathcal{E}_M}  \propto B$, where $B$ is the magnetic field strength, 
the damping stage of dynamo is characterized by a linear-in-time growth of $B$.

From Eqs. \eqref{eq: mk} and \eqref{eq: inmke}, we find 
\begin{equation}
     \frac{d \ln \mathcal{E}_M}{dt} \propto \Gamma_d.
\end{equation}
Here $\Gamma_d \propto \mathcal{E}_M^{-\frac{1}{2}}$ according to Eqs. \eqref{eq: gamd} and \eqref{eq: lde}, 
which results from both the equalization $\Gamma_d = \omega_\text{IN}$ at $l_d$ and the Kolmogorov scaling of turbulence. 
Therefore, we have $\sqrt{ \mathcal{E}_M}  \propto t$.

Furthermore, after inserting Eq. \eqref{eq: emdt} into Eq. \eqref{eq: lde}, we can also derive the time evolution of $l_d$, 
\begin{equation}\label{eq: ldevlt}
    l_d = \Big(l_{d1}^\frac{2}{3} + \frac{3}{23} L^{-\frac{1}{3}} V_L (t-t_1)\Big)^\frac{3}{2},
\end{equation}
with $l_{d1}$ at $t=t_1$.
If the damping stage can proceed until $l_d$ increases up to $L$, {\it Condition (1)} (Eq. \eqref{eq: conlen}) should be satisfied at $L$, that is, 
\begin{equation}\label{eq: edinl}
     \frac{2 L}{ \mathcal{C} V_L} < 1. 
\end{equation}
Compared with the general form in Eq. \eqref{eq: conlen}, the above condition requires a further smaller ionization fraction 
so that even the largest eddy-turnover time is still smaller than the neutral-ion collisional time. 
With neutrals decoupled from the {dynamo action}
on all length scales from the initial $l_d$ up to $L$, it ensures that 
the dynamo remains in the {kinematic damping stage}, and 
the unsaturated magnetic energy at the end of dynamo mainly comes from the turbulent energy carried by ions.

When $l_d = L$, the corresponding time is (Eq. \eqref{eq: ldevlt}), 
\begin{equation}
    t(l_d = L) = t_1 + \frac{23}{3}  L^\frac{1}{3} V_L^{-1} (L^\frac{2}{3} - l_{d1}^\frac{2}{3}) .
\end{equation}
Given $l_{d1} \ll L$, the entire damping stage of dynamo lasts for around $7.7$ times the largest eddy-turnover time.
The magnetic energy reached at $l_d = L$ is (Eq. \eqref{eq: lde}), 
\begin{equation}\label{eq: fine}
     \mathcal{E}_{M} (l_d=L) = \mathcal{C}^{-1} L V_L .
\end{equation}
In the kinematic damping stage, there is 
\begin{equation}
     \mathcal{E}_{M} (l_d=L)  < \frac{1}{2} V_L^2, 
\end{equation}
which naturally recovers the condition in Eq. \eqref{eq: edinl}.

\begin{figure*}[htbp]
\centering   

\subfigure[Dissipation-free stage]{
   \includegraphics[width=8.5cm]{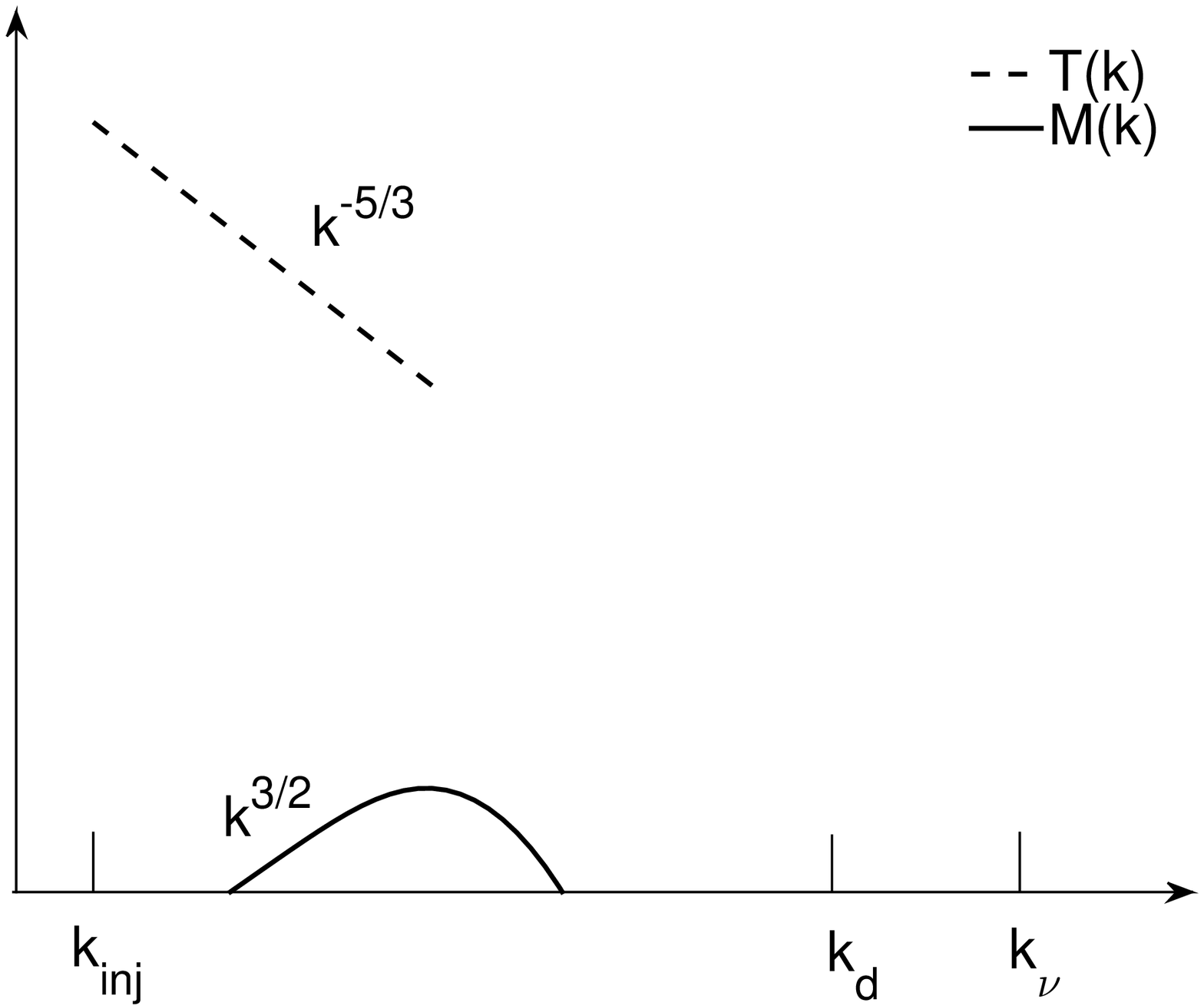}\label{fig: sep}}
\subfigure[Damping stage]{
   \includegraphics[width=8.5cm]{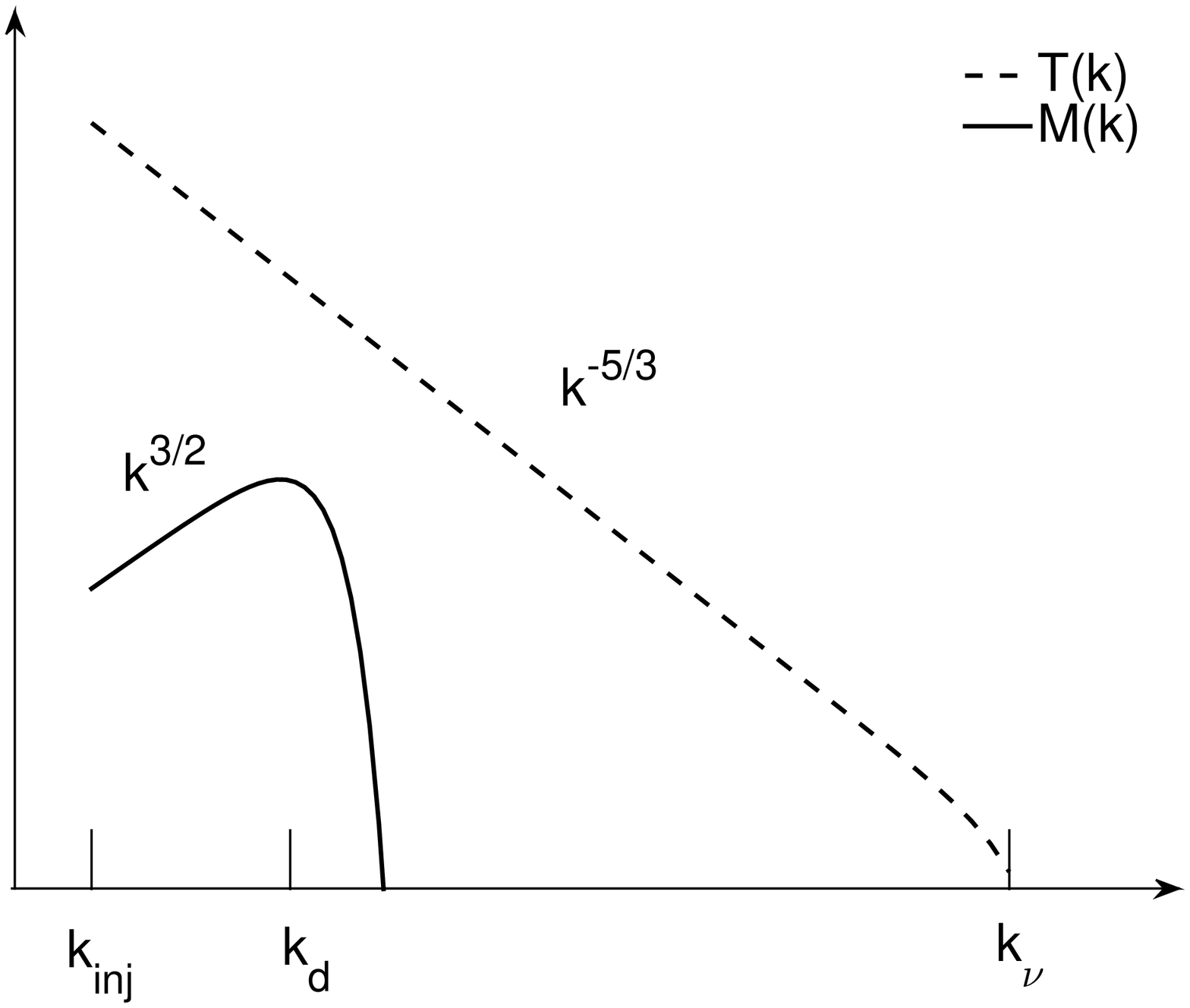}\label{fig: skd}}

\caption{Sketches of the magnetic energy spectrum $M(k)$ and the turbulent kinetic energy spectrum $T(k)$ for the dissipation-free and 
damping stages of dynamo, respectively.
$k_\text{inj}$, $k_d$, $k_\nu$ are the wavenumbers corresponding to $L$, $l_d$, and $l_\nu$.}
\label{fig: ske}
\end{figure*}

\begin{figure*}[htbp]
\centering   
   \subfigure[]{  
   \includegraphics[width=8.5cm]{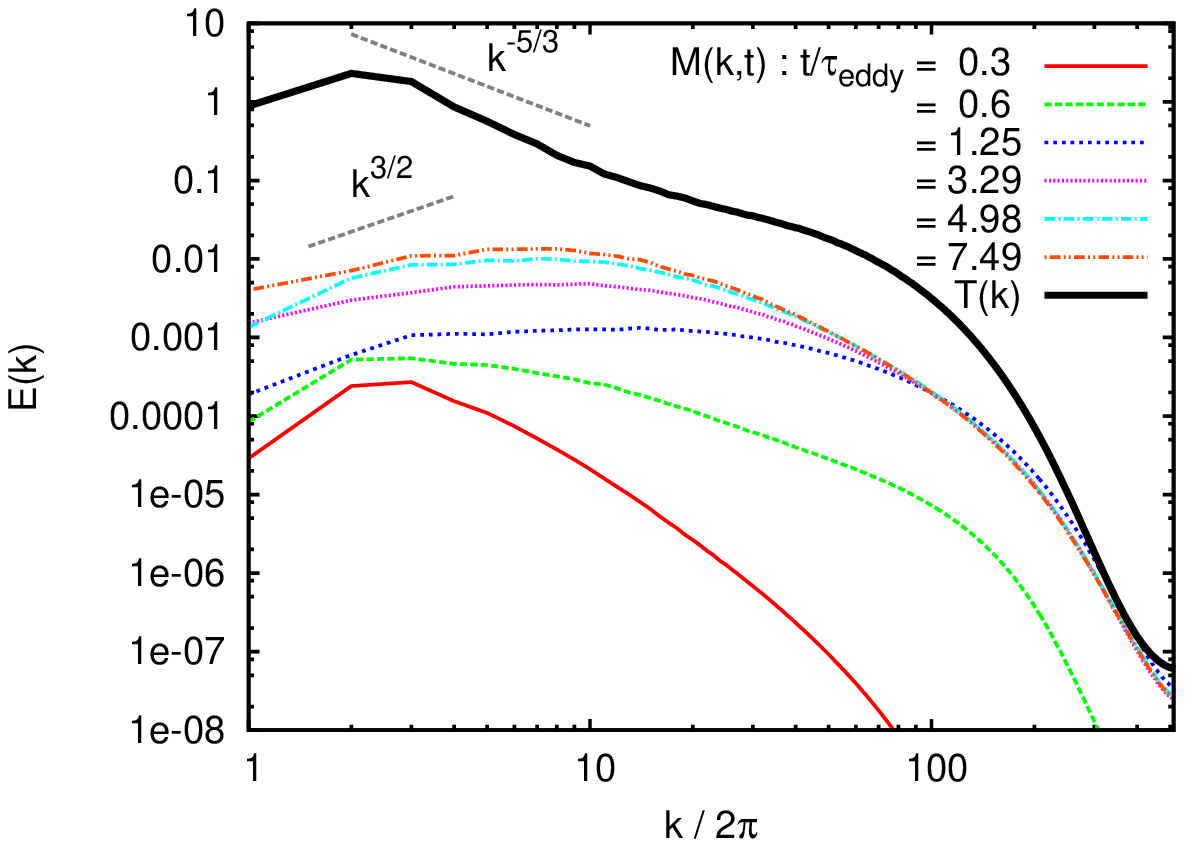}\label{fig: numev}}
   \subfigure[]{
   \includegraphics[width=8.5cm]{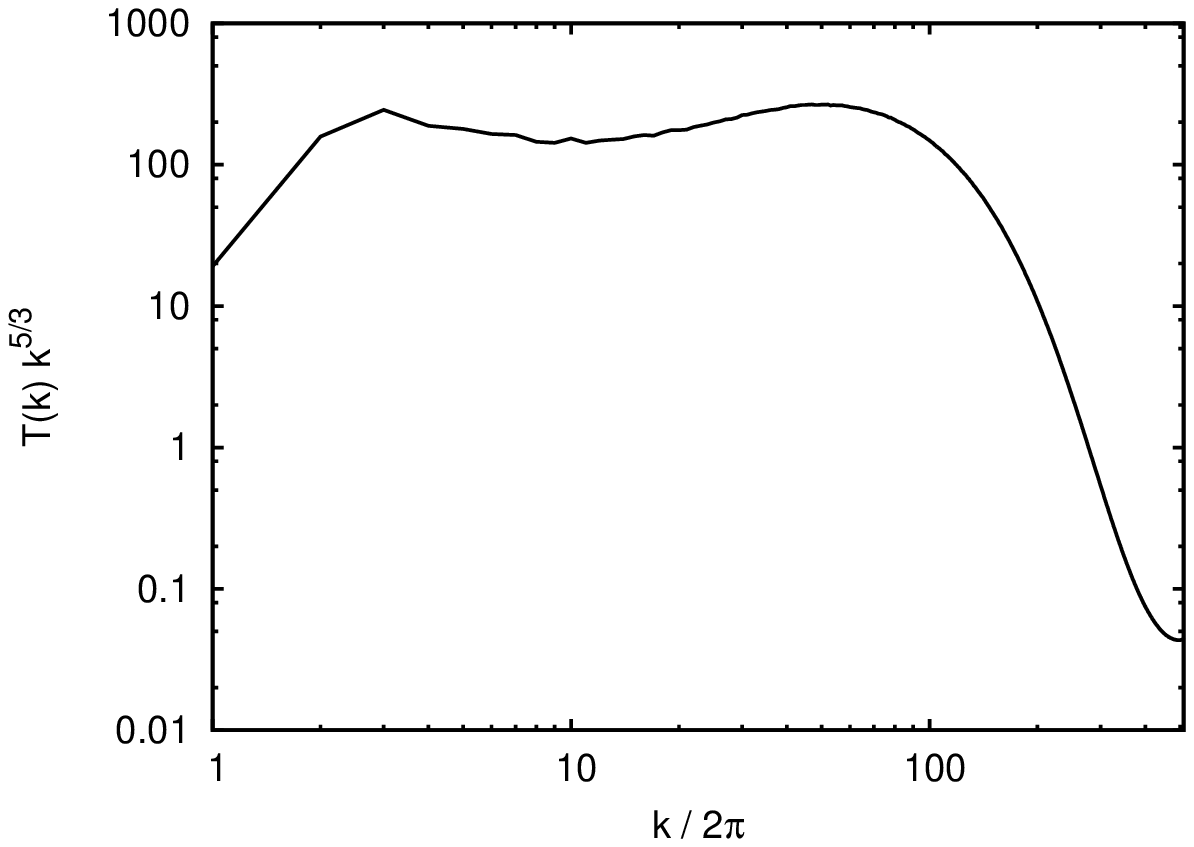}\label{fig: botn}}
   
\caption{(a) Numerically measured $M(k,t)$ at different times. 
$T(k)$ is the steady turbulent energy spectrum. 
The short dashed lines indicate the spectral scalings of the Kazantsev spectrum and Kolmogorov spectrum. 
(b) {Compensated turbulent energy spectrum.}}
\end{figure*}

\begin{figure*}[htbp]
\centering   
   \subfigure[]{
   \includegraphics[width=8.5cm]{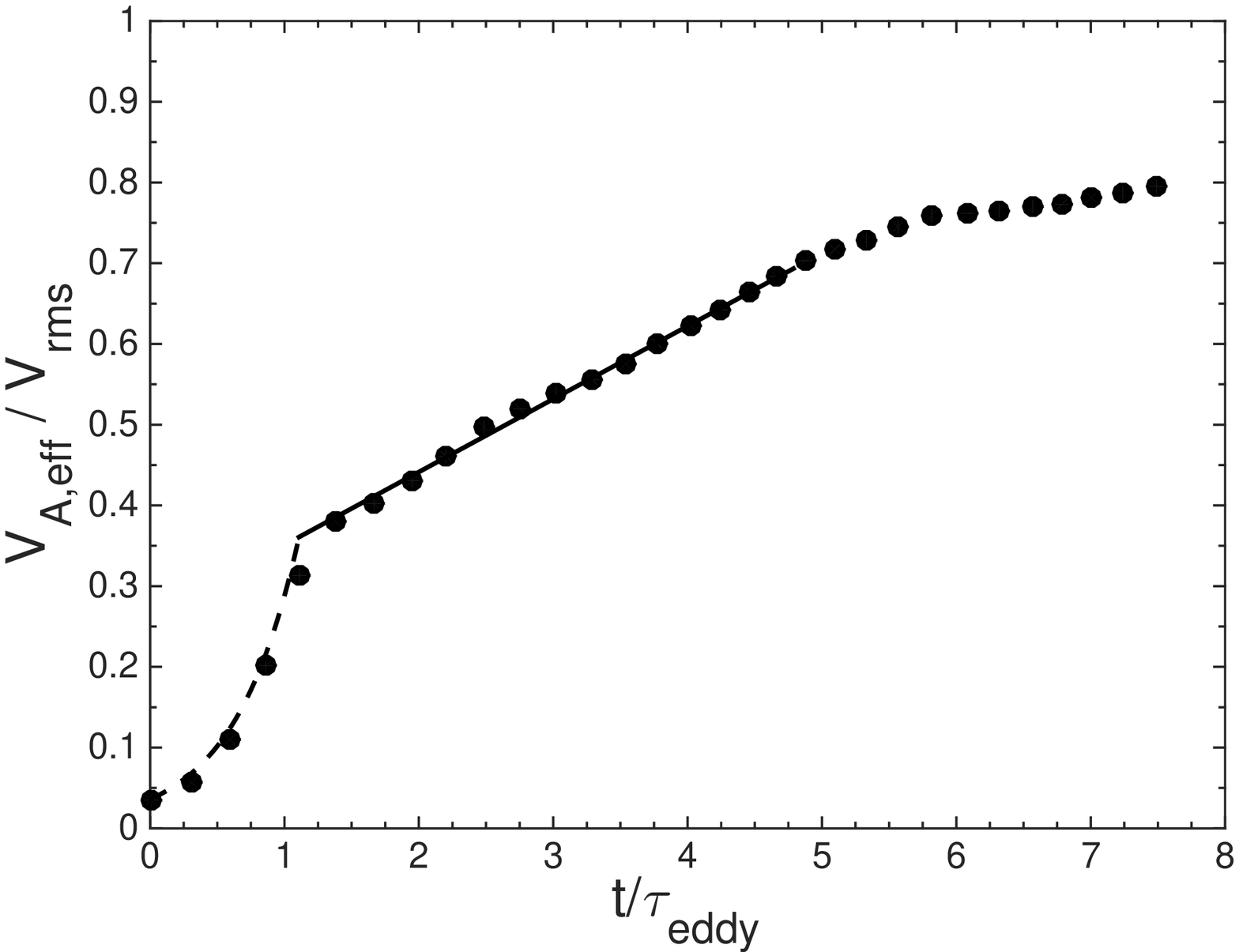}\label{fig: edli}}
   \subfigure[]{
   \includegraphics[width=8.5cm]{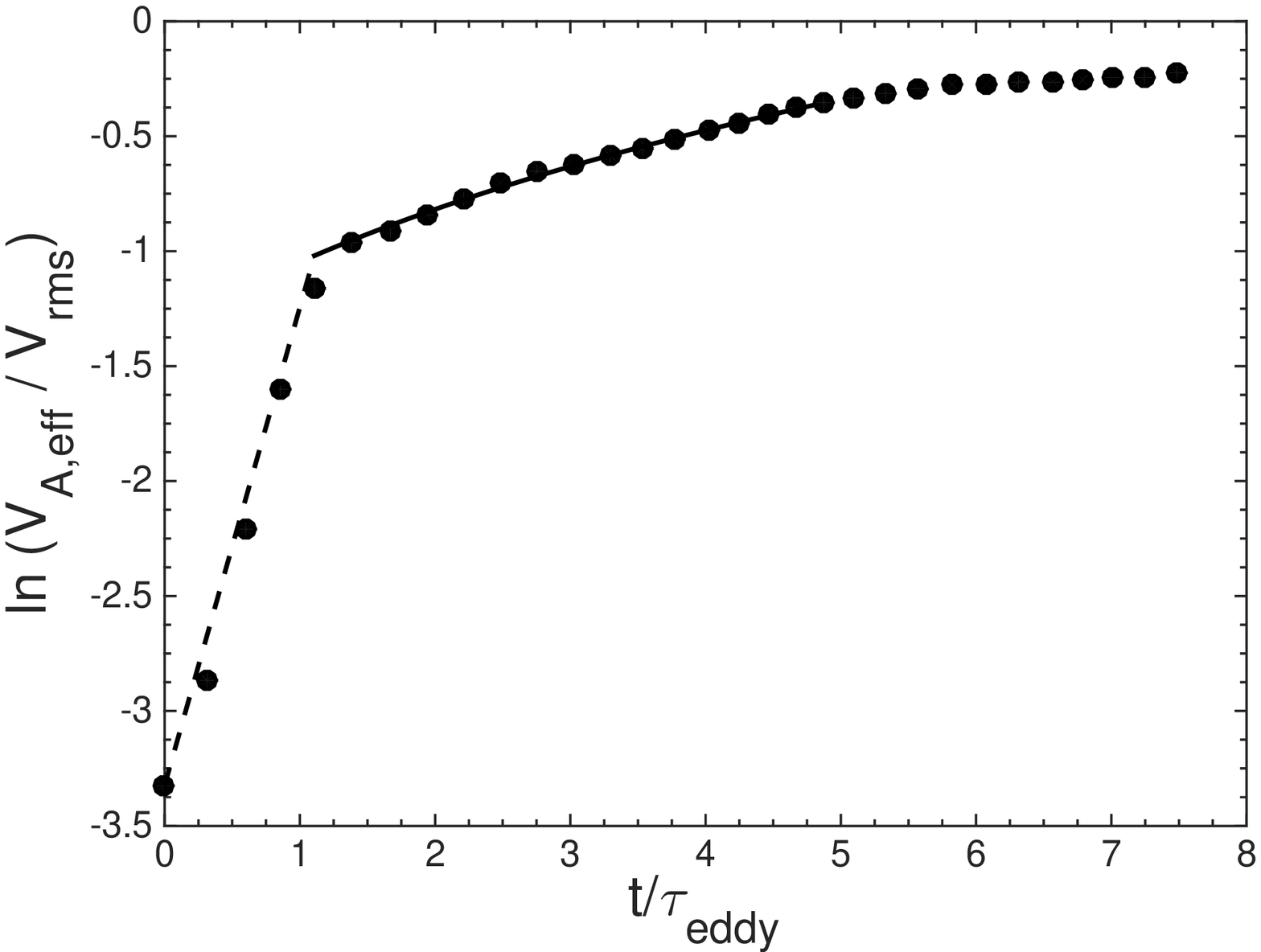}\label{fig: loged}}

\caption{ (a) Time evolution of the numerically measured $V_{A,\text{eff}}$ (normalized by $V_\text{rms}$, filled circles) 
in comparison with our theoretical prediction, where the dashed and solid lines represent the 
dissipation-free (Eq. \eqref{eq: va1k}) and damping (Eq. \eqref{eq: va2d}) stages of dynamo, respectively.
{(b) Same as (a) but for the logarithm of $V_{A,\text{eff}} / V_\text{rms}$. }}
\label{fig: ed}
\end{figure*}

\section{Numerical test of the damping stage of dynamo with a two-fluid simulation }
\label{sec: num}

To numerically test the above theory for the damping stage of dynamo,
we perform a 3D two-fluid dynamo simulation by using the RIEMANN code
\citep{Bal98a,Bal98b,Bal04,Bal10,Bal12,BS99a,BS99b}.
The neutral and ionized fluids are separately treated with the isothermal Euler equations
and isothermal MHD equations, respectively. 
Their coupling is described by the ion-neutral friction term, which is introduced using an operator-split method
\citep{TB08,Til12}. 
We solve the following equations
\citep{Dra86}
using the above mentioned code:
\begin{equation}
\begin{aligned}
&\frac{\partial \rho_i}{\partial t} + \nabla \cdot (\rho_i \boldsymbol{v}_i) = 0,  \\
&\frac{\partial \boldsymbol{v}_i}{\partial t} + (\boldsymbol{v}_i \cdot \nabla)\boldsymbol{v}_i =
     -c_s^2 \nabla \ln \rho_i - \frac{1}{4\pi}\boldsymbol{B}\times(\nabla \times \boldsymbol{B})  \\
&  ~~~~~~~~~~~~~~~~~~~~~~~~~~~~~~~~   - \gamma_d \rho_n (\boldsymbol{v}_i - \boldsymbol{v}_n),  \\
&\frac{\partial \boldsymbol{B}}{\partial t} = \nabla \times ( \boldsymbol{v}_i \times \boldsymbol{B}), \\
&\frac{\partial \rho_n}{\partial t} + \nabla \cdot (\rho_n \boldsymbol{v}_n) = 0,  \\
&\frac{\partial \boldsymbol{v}_n}{\partial t} + (\boldsymbol{v}_n \cdot \nabla)\boldsymbol{v}_v =
      -c_s^2 \nabla \ln \rho_n - \gamma_d \rho_i (\boldsymbol{v}_n - \boldsymbol{v}_i), 
\end{aligned}
\end{equation}
where $\boldsymbol{v_i}$ and $\boldsymbol{v_n}$ are the velocities of the ionized and neutral fluids, 
and $\boldsymbol{B}$ is the magnetic field.
As the time step is restricted by the Alfv\'{e}n time step for ions, a two-fluid simulation at a low ionization fraction
is computationally very expensive.
To reduce the computational cost,
a ``heavy ion approximation" 
(HIA, \citealt{Ois06,Li08})
with artificially decreased ion Alfv\'{e}n speed and increased ionization fraction 
is frequently adopted. 
However, 
\citet{TilBal10}
showed that the HIA can unphysically affect the dissipation characteristics of magnetic fluctuations. 
It is also possible to numerically investigate the partially ionized magnetized fluids using
the single-fluid treatment by adding an additional diffusive term in the induction equation
\citep{Os06,Os07}.
However, this approach is unable to capture 
the two-fluid effect in the weakly coupled regime
\citep{Bals96,Xuc16},
which is of key importance to study the damping stage of dynamo considered here.
Therefore, we perform a full two-fluid simulation with realistic ion masses to obtain reliable numerical measurements.

\subsection{Simulation setup}

We set initially uniform densities of both ions and neutrals, with the neutral density equal to unity. 
{The ions and neutrals have 
molecular weights as $\mu_i = 29$ amu (corresponding to HCO+) and $\mu_n = 2.3$ amu (corresponding to 
H$_2$ and He), respectively, 
as the mean molecular mass of ions and neutrals in molecular clouds
\citep{Shu92,Bals96,TilBal10,Mey14,Burk15}.}
The RIEMANN code has been used to simulate two-fluid magnetized turbulence with an ionization fraction as low as $10^{-6}$
\citep{TB08,TilBal10}.
Here we choose a value of $10^{-4}$. 
The initial seed magnetic field for dynamo amplification is uniform
\citep{CVB09}
and aligned along the x-direction. 
We drive hydrodynamical turbulence in this
initial setup.   
The hydrodynamic turbulence is forced via driving random Gaussian fluctuations in Fourier space, with the driving scale peaked at $k/2\pi=2$ and 
spanning $1 \leq k/2\pi \leq 4$, 
and an rms velocity of $0.2$ times the sound speed.
{The turbulence becomes statistically steady after around three turnover times of the largest eddy.} 
We continuously drive the turbulence in both ions and neutrals to maintain a constant turbulent energy 
{and a constant rms velocity}
throughout the simulation.
The turbulent energy cascades toward smaller scales and dissipates at the numerical dissipation scale. 
To ensure a clear separation between the driving scale of turbulence, 
the ion-neutral collisional damping scale that increases with time, 
and the numerical dissipation scale of turbulence, our simulation has a high resolution of $1024^3$ mesh points. 
It is performed in a computational domain given by $[0,1]\times[0,1]\times[0,1]$.

Table \ref{tab: par} lists
the numerical resolution ({mesh points}), 
{the injection scale ({mesh points}) where most turbulent energy is injected,}
the ratio between $\rho_i$ and $\rho_n$, 
the rms velocity, 
the sound speed, 
{the initial Alfv\'{e}n Mach number 
$M_{Ai0} = V_\text{rms} / V_{Ai0}$ of the ionized fluid, 
where $V_{Ai0} = B_0 /\sqrt{4\pi \rho_i}$ is the initial Alfv\'{e}n speed in terms of the initial magnetic field strength $B_0$ 
and ion density $\rho_i$,
the ratio between $\nu_{ni}$ and the eddy-turnover rate $V_\text{rms} / L $ at $L$.
The large value of $M_{Ai0}$ shows that the initial magnetic energy contained in the ionized fluid 
is much smaller than the turbulent energy. }

 With our focus on the damping stage of dynamo, we designate the values of above parameters in the simulation to satisfy the physical conditions 
presented in Section \ref{sec: damthe}.

\begin{table}[t]
\renewcommand\arraystretch{1.5}
\centering
\begin{threeparttable}
\caption[]{Simulation parameters}\label{tab: par} 
  \begin{tabular}{ccccccc}
     \toprule
       $R$                &    $L$                   &    $\rho_i/\rho_n$              & $V_\text{rms}$  & $c_s$  &    $M_{Ai0}$     &   $\frac{\nu_{ni}}{V_\text{rms}/L}$    \\
       $1024^3$       &    $512$               &     $1.26\times10^{-3}$      &  $0.2$               &  $1$     &     $17.7$          &   $0.08$       \\
    \bottomrule
    \end{tabular}
 \end{threeparttable}
\end{table}

{\it Condition (1)}:

To ensure that the dynamo stage can proceed until $l_d = L$, 
we have (Eq. \eqref{eq: parac}, Eq. \eqref{eq: edinl})
\begin{equation}\label{eq: con1}
    \frac{2 L}{\mathcal{C} V_L }  
    \approx 0.48 < 1,
\end{equation}
where the values in Table \ref{tab: par} are adopted and we take $V_L = V_\text{rms}$. 
{It shows that due to the low ionization fraction, neutrals are decoupled from the dynamo action on all length scales. }

{\it Condition (2)}:
 
We note that {\it Condition (2)} is naturally fulfilled given {\it Condition (1)} (see Section \ref{ssec: con}).
{Due to the low ionization fraction and strong ion-neutral collisional damping, there is no equipartition between magnetic and turbulent energies on all length scales. 
At the end of damping stage at $l_d = L$,
the unsaturated magnetic energy mainly contained in ions is smaller than the turbulent energy at $L$.}

{\it Condition (3)}:

The initial magnetic energy contained in ions is 
\begin{equation}
    \mathcal{E}_{M0}  = \frac{1}{2} V_{Ai0}^2 .
\end{equation}
We rewrite Eq. \eqref{eq: con2} in a dimensionless form and find 
\begin{equation}
    \frac{\mathcal{E}_{M0}}{\mathcal{C}^{-1} \nu_n} = \frac{1}{6} \frac{1}{M_{Ai0}^2}   \frac{V_L/L}{\nu_{ni}} \Big(\frac{L}{l_\nu}\Big)^\frac{4}{3} 
    \approx  1.3 > 1,
\end{equation}
where the viscous scale is
\begin{equation}
     l_\nu = L^\frac{1}{4} V_L^{-\frac{3}{4}} \nu_n^\frac{3}{4}.
\end{equation}
{In our simulation, 
$l_\nu$ is determined by the numerical dissipation scale, which is on the order of $10$ mesh points.}
With the growth of $\mathcal{E}_M$, we have the ion-neutral collisional damping as the dominant damping effect and 
{\it Condition (3)} is satisfied.

\subsection{Comparison between theoretical predictions and numerical measurements}
\label{ssec: com}

{At an early time of the simulation, 
before the turbulent energy spectrum is fully developed, 
due to the turbulent energy cascade from large to small scales, 
the dynamo stretching scale, which determines the peak scale of magnetic energy spectrum, 
shifts toward smaller scales. 
The initial weak magnetic field leads to the initially weak ion-neutral collisional damping effect.} Thus 
the dynamo is in the dissipation-free regime, which is characterized by an exponential growth of magnetic energy. 
A Kazantsev magnetic energy spectrum on scales larger than the peak scale is expected, 
as seen earlier in one-fluid dynamo simulations 
(e.g., \citealt{Hau04,Bran05}).

With the growth of magnetic energy, the ion-neutral collisional damping becomes important, 
so that the magnetic energy spectrum peaks and is also damped at $l_d$. 
The dynamo enters the damping stage. 
As analyzed in Section \ref{ssec: the}, we expect that 
the magnetic field strength grows linearly with time, and the spectral peak at $l_d$ moves toward larger 
scales.

In Fig. \ref{fig: ske}, we illustrate $M(k)$ in both the dissipation-free stage and the damping stage. 
As a comparison, 
the numerically measured $M(k,t)$ at different times are presented in Fig. \ref{fig: numev}. 
As expected, the spectral peak of $M(k,t)$ indeed first shifts to smaller scales and then back to larger scales. 
The ascending spectral form on large scales is also consistent with the Kazantsev spectrum $\sim k^{3/2}$. 
Besides, we also present the stationary and fully developed
turbulent energy spectrum $T(K)$, which is expected to follow the Kolmogorov spectrum $k^{-5/3}$.
The appearance of a bottleneck effect with a pileup of energy 
\citep{Fal94}
is observed on small scales of the inertial range, 
{which can be more clearly seen in the compensated turbulent energy spectrum in Fig. \ref{fig: botn}.}

1. Dissipation-free stage of dynamo

When the ion-neutral collisional damping effect is weak, the dynamo stretching leads to an exponential growth of magnetic energy, 
\begin{equation}\label{eq: exp}
     \mathcal{E}_M = \mathcal{E}_{M0} \exp (2 \Gamma_p t).
\end{equation}
The dynamo growth rate $\Gamma_p$ corresponds to the eddy-turnover rate at the peak scale $l_p$ of 
$M(k,t)$. 
As the spectral peak shifts toward smaller scales, 
$\Gamma_p$ increases with time. 

To compare with the numerical result, we rewrite Eq. \eqref{eq: exp} in the form, 
\begin{equation}\label{eq: va1k}
      \frac{V_{A1,\text{eff}}}{V_L} = \frac{V_{A0,\text{eff}}}{V_L}  \exp \Big(  \Big(\frac{L}{l_p}\Big)^\frac{2}{3}   \frac{t}{\tau_\text{eddy}}  \Big),  
\end{equation}
where $ \tau_\text{eddy} =  L/V_L$ is the turnover time of the largest eddy at $L$, and 
$V_{A0,\text{eff}}$ and $V_{A1,\text{eff}}$ are the effective Alfv\'{e}n speeds in terms of the effective density $\rho_\text{eff}$
at the beginning and the end of the dissipation-free stage. 
We define $\rho_\text{eff}$ as 
\begin{equation}\label{eq: efdid}
   \rho_\text{eff} =  \eta_c \rho_i, 
\end{equation}
where $\eta_c$ is the coupling coefficient, as an indicator of the coupling degree between ions and neutrals. 
When ions and neutrals are strongly coupled together, there is $\eta_c = \rho/\rho_i$ 
and $V_{A, \text{eff}} = V_{A, \text{tot}}$, where $V_{A, \text{tot}}$ is the Alfv\'{e}n speed in terms of the total density. 
When ions and neutrals are decoupled from each other, 
we have $\eta_c = 1$ and $V_{A, \text{eff}} = V_{Ai}$. 
Here we are concerned with the weak coupling regime with $\eta_c \gtrsim 1$, 
where neutrals are decoupled from ions but ions are still coupled with neutrals. 
The exact value of $\eta_c$ will be determined numerically.

By adopting the values in Table \ref{tab: par}, 
we present the above theoretical calculation (Eq. \eqref{eq: va1k}) in comparison with the numerical result in Fig. \ref{fig: ed}.
Approximately, we use a constant value of $l_p \sim L/3$ as an estimate of the evolving $l_p$ 
and find 
\begin{equation}\label{eq: f1end}
    \frac{V_{A1,\text{eff}}}{V_L} \approx  0.36
\end{equation}
at the end of the dissipation-free stage at $t=t_1 = 1.1 \tau_\text{eddy}$.

{We would like to stress here that the growth of magnetic energy during the dissipation-free stage 
indeed enhances the damping effect, 
but the key and necessary condition for the damping stage to arise is a sufficiently small ionization fraction, 
i.e., {\it Condition (1)}
(Eq. \eqref{eq: con1}).}

2. Damping stage of dynamo

We rewrite the evolution law of magnetic energy in the damping stage of dynamo given by 
Eq. \eqref{eq: emdt} in a dimensionless form 
\begin{equation}\label{eq: va2d}
     \frac{V_{A2,\text{eff}}}{V_L} =  \frac{V_{A1,\text{eff}}}{V_L} + \frac{3 \sqrt{2}}{23} \Big( \frac{3 \nu_{ni} L}{V_L} \Big)^\frac{1}{2} \frac{(t - t_1)}{\tau_\text{eddy}} .
\end{equation}
With the values of parameters in Table \ref{tab: par} used, 
the theoretical calculation is displayed in Fig. \ref{fig: ed}. 
By comparing with the numerical measurement, 
we also found $\eta_c \approx 2.45$. 
With $\eta_c$ being of the order of a few, $\rho_\text{eff}$ is close to $\rho_i$ (Eq. \eqref{eq: efdid}). 
It shows that the growing magnetic energy mainly comes from the turbulent energy contained in ions 
in the weak coupling regime, as discussed in Section \ref{ssec: con}.

At the end of the damping stage, the theoretical expectation in Eq. \eqref{eq: fine} yields 
\begin{equation}\label{eq: f2end}
       \frac{V_{A2,\text{eff}}}{V_L} = \Big(6 \frac{\nu_{ni} L } {V_L}\Big)^\frac{1}{2}  
       = 0.69 < 1.
\end{equation}
The corresponding time is (Eqs. \eqref{eq: f1end}, \eqref{eq: va2d}, and \eqref{eq: f2end}),
\begin{equation}\label{eq: endamt}
  t_2  = t_1 + 3.7 \tau_\text{eddy} = 4.8 \tau_\text{eddy} .
\end{equation}  
We see in Fig. \ref{fig: ed} that the damping stage observed in the numerical simulation is slightly more extended than the above prediction, 
but the dynamo growth ceases soon after $t = t_2$.

\begin{figure*}[htbp]
\centering   
   \subfigure[]{
   \includegraphics[width=7.5cm]{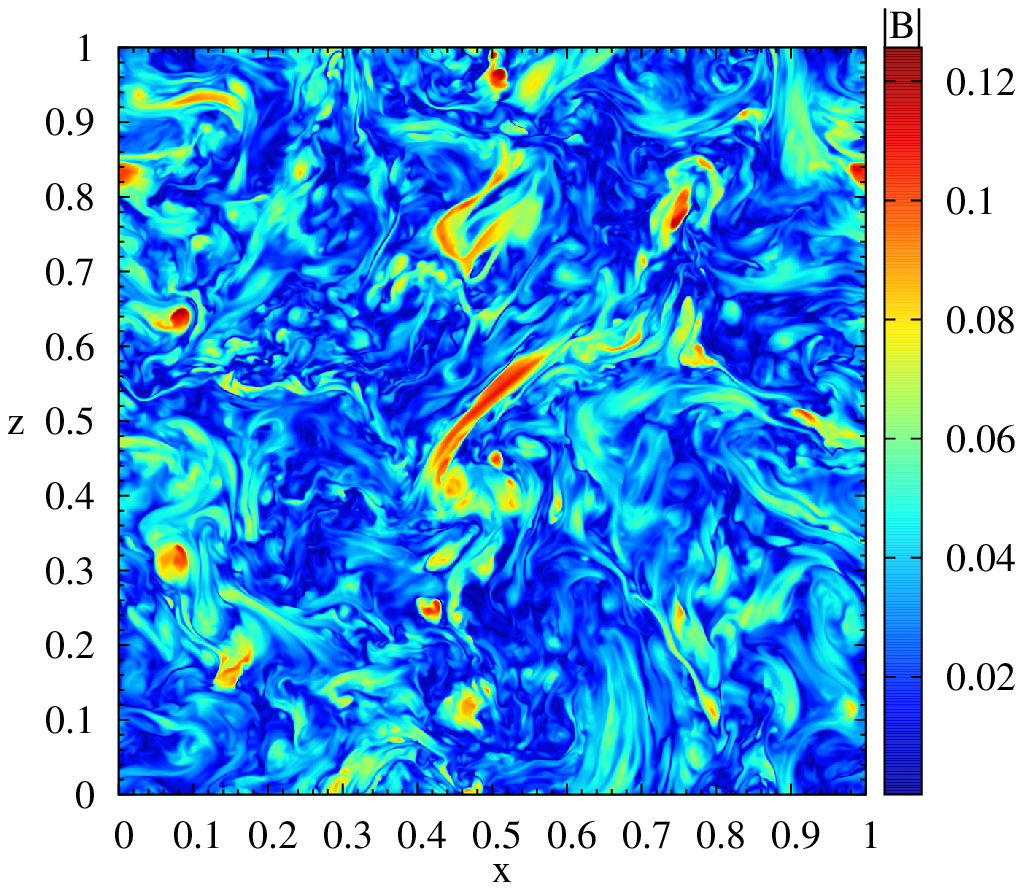}\label{fig: magstra}}
   \subfigure[]{
   \includegraphics[width=7.5cm]{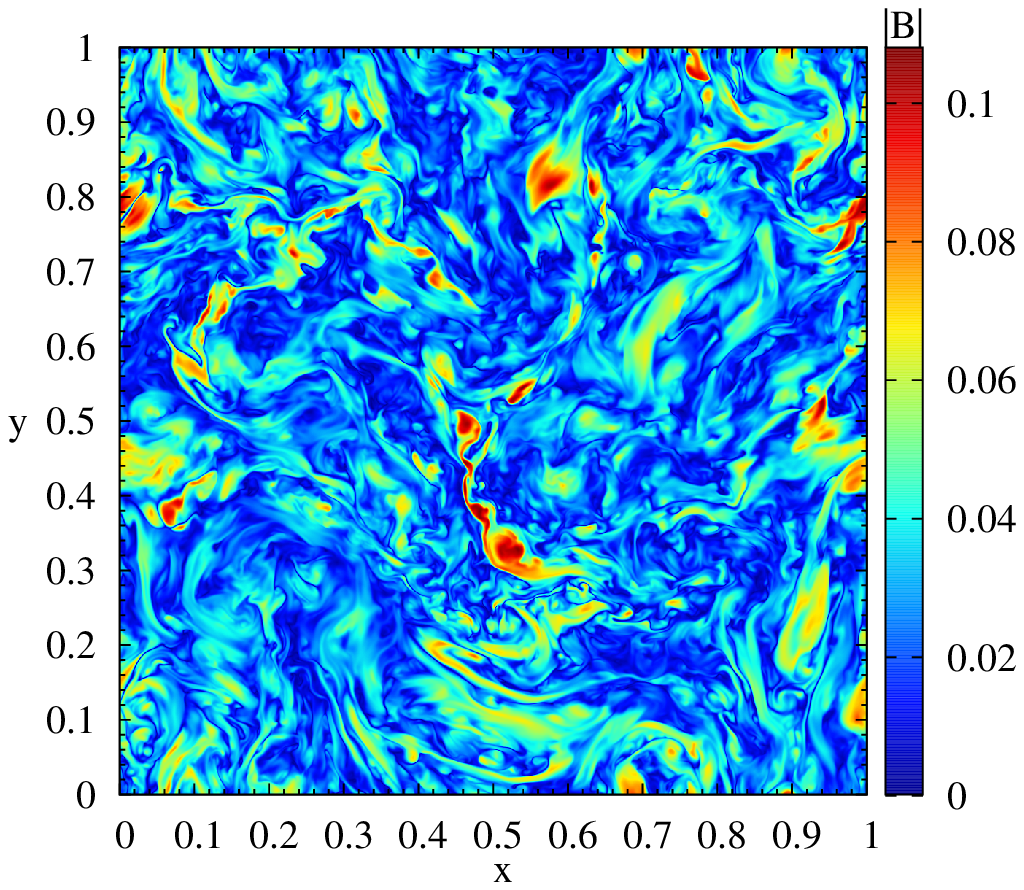}\label{fig: magstrb}}
   \subfigure[]{
   \includegraphics[width=7.5cm]{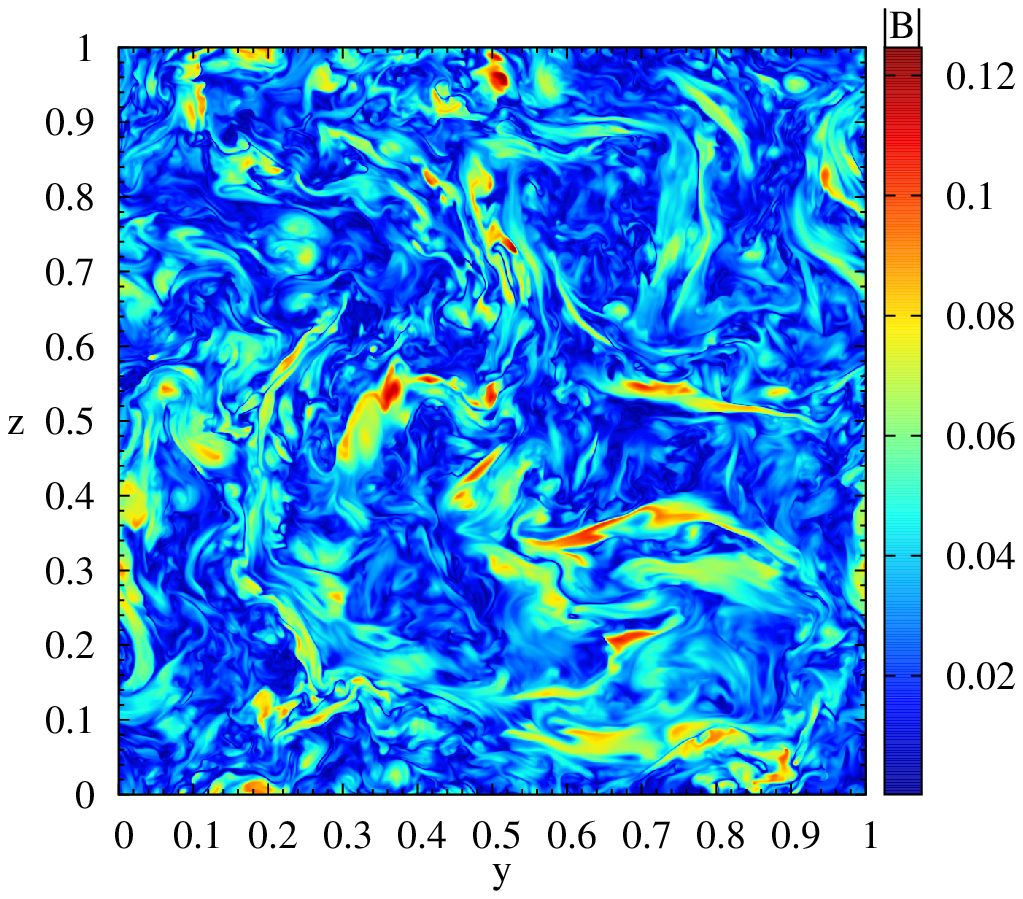}\label{fig: magstrc}}
\caption{A 2D cross section through the middle of the computational domain of the numerically measured magnetic field strength
{in the (a) xz plane, (b) xy plane, and (c) yz plane,} 
corresponding to $M(k,t)$ at $t=7.49~\tau_\text{eddy}$ in Fig. \ref{fig: numev}.}
\label{eq: ms2dldst}
\end{figure*}

Moreover, the time evolution of $l_d$ in the damping stage is (Eq. \eqref{eq: ldevlt})
\begin{equation}\label{eq: ldevtn}
    \frac{l_d}{L} = \bigg[\bigg(\frac{l_{d1}}{L}\bigg)^\frac{2}{3} + \frac{3}{23}   \frac{(t-t_1)}{\tau_\text{eddy}}\bigg]^\frac{3}{2}.
\end{equation}
Starting from (Eqs. \eqref{eq: lde} and \eqref{eq: f1end})
\begin{equation}
   \frac{l_{d1}}{L}  =\bigg[\frac{V_L}{6 \nu_{ni} L} \frac{V_{A1,\text{eff}}^2}{V_L^2}\bigg]^\frac{3}{4} = 0.38,
\end{equation}
we see that $l_d$ reaches $L$ at $t = t_2$ (Eqs. \eqref{eq: endamt} and \eqref{eq: ldevtn}). 
Fig. \ref{eq: ms2dldst} displays the 2D magnetic field structure measured at the end of the simulation, 
which is dominated by large-scale magnetic field fluctuations.   
It confirms that the magnetic field resulting from the damping stage of dynamo 
has a characteristic length scale comparable to $L$ in our simulation.

\section{Physical conditions in the ISM for the damping stage of dynamo}
\label{sec: phsim}

As illustrative examples for the applications of the above dynamo theory, here we examine the physical conditions in the partially ionized ISM for 
the damping stage of dynamo. 
Table \ref{Tab: ism} lists the typical parameters of the warm neutral medium (WNM), the cold neutral medium (CNM), molecular clouds (MC) 
and dense cores in molecular clouds (DC), 
where $n_H$ and $n_e$ are 
number densities of the atomic hydrogen and electrons, and $T$ is the temperature. 
Their values are taken from 
\citet{Dra98}.
Besides, we assume $m_i=m_n=m_\text{H}$ as the masses of ions and neutrals in WNM and CNM,
and $m_i=29 m_\text{H}$, $m_n=2.3 m_\text{H}$ in MC and DC
\citep{Shu92}, 
where $m_\text{H}$ is the hydrogen atomic mass. 
We also have $\nu_n = v_\text{th}/ (n_n \sigma_{nn})$, with the neutral thermal speed $v_\text{th}$, the neutral number density $n_n$, 
and the cross-section of a neutral-neutral collision $\sigma_{nn} \sim 10^{-14} ~\text{cm}^2$
\citep{VrKr13}. 
The drag coefficient is 
{$\gamma_d=5.5\times10^{14}$cm$^3$g$^{-1}$s$^{-1}$ in WNM and CNM, 
and 
$\gamma_d=3.5\times10^{13}$cm$^3$g$^{-1}$s$^{-1}$ in MC and DC
\citep{Drai83,Shu92}.}
We next analyze the turbulent dynamo induced by 
(a) the globally driven interstellar turbulence and 
(b) the locally excited turbulence in supernova remnants (SNRs).

\begin{table}[h]
\renewcommand\arraystretch{1.3}
\centering
\begin{threeparttable}
\caption[]{Turbulent dynamo in the partially ionized ISM
}\label{Tab: ism} 
  \begin{tabular}{ccccc}
      \toprule
 & WNM & CNM & MC & DC \\
      \midrule
$n_\text{H}$[cm$^{-3}$]  & $0.4$ & $30$ & $300$ & $10^4$  \\
$n_e/n_\text{H}$  & $0.1$ & $10^{-3}$ & $10^{-4}$ & $10^{-6}$  \\
$T$[K]  & $6000$ & $100$ & $20$ & $10$  \\
\cline{1-5}
           \multicolumn{5}{c}{Interstellar turbulence}    \\
\cline{1-5}
$l_{d,\text{cr}}$ [pc]  &   -  & -  & $6.3\times10^{-6}$ & $3.3\times10^{-5}$  \\
$\tau_\text{dam}$[kyr]   & - & -                    & $0.4$                     &  $2.3$    \\
$B_\text{dam}$ [$\mu$ G] & - & - & $0.5$  & $4.7$    \\
$\tau_\text{non}$[kyr]    &  $1.9\times10^4$  & $1.9\times10^4$  & $1.9\times10^4$  & $1.9\times10^4$  \\
$B_\text{non}$ [$\mu$ G]  & $3.0$  & $25.1$   & $79.5$  & $458.1$ \\
\cline{1-5}
           \multicolumn{5}{c}{Preshock turbulence}   \\
\cline{1-5}
$l_{d,\text{cr}}$ [pc]  & \multicolumn{4}{c}{$0.1$}   \\
$\tau_\text{dam}$[kyr]   &   \multicolumn{4}{c}{$0.75$}  \\
$B_\text{dam}$ [$\mu$ G]  & $79.1$  & $56.6$  & $415.2$  & $138.2$ \\
 \bottomrule
    \end{tabular}
 \end{threeparttable}
\end{table}

(a) Interstellar turbulence 

\begin{figure*}[htbp]
\centering   
   \subfigure[Interstellar turbulence]{  
   \includegraphics[width=8.5cm]{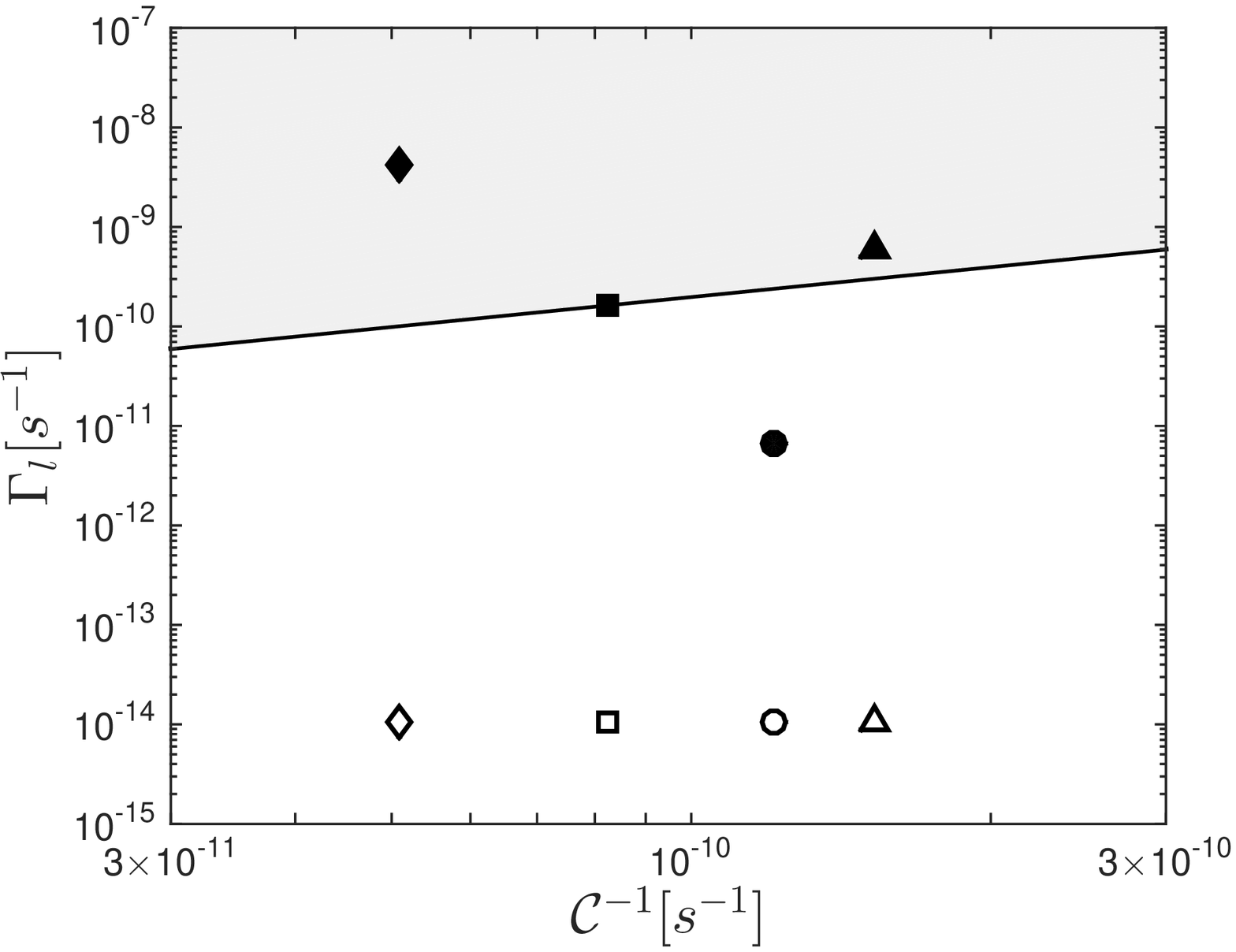}\label{fig: ism}}
   \subfigure[Preshock turbulence]{
   \includegraphics[width=8.5cm]{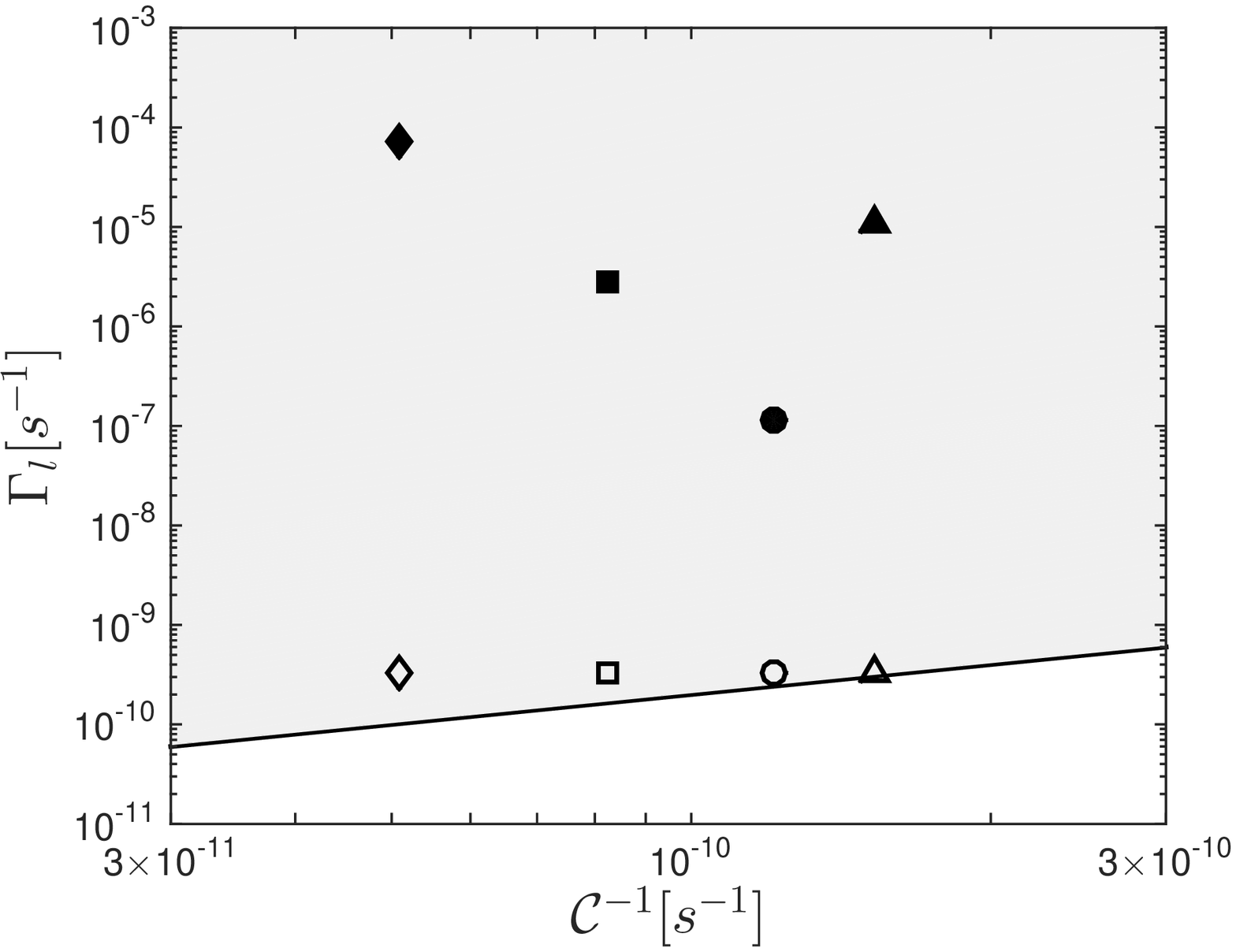}\label{fig: isms}}
\caption{The shaded region shows the parameter space for the appearance 
of damping stage of dynamo. The symbols represent the values for 
WNM (circle), CNM (square), MC (triangle), and DC (diamond). 
Filled and open symbols correspond to {$\Gamma_\nu$ and $\Gamma_L$}, respectively. }
\label{fig: ismts}
\end{figure*}

We consider that the interstellar turbulence driven by 
supernova explosions has a typical driving condition
\citep{Spi78},
\begin{equation}
       L=30\, \text{pc},   ~V_L=10 \,\text{km s}^{-1}.
\end{equation}
As a result of turbulent energy cascade, the interstellar turbulence extends from $L$ to $l_\nu$.
{Here we assume that the initial seed magnetic field is sufficiently weak, so that} 
the turbulent motions on all length scales can contribute to the dynamo growth. 
In partially ionized phases, 
to examine {\it Condition (1)},
we calculate the {dynamo stretching rate $\Gamma_\nu$ of $l_\nu$-scale eddies and 
$\Gamma_L$ of $L$-scale eddies in comparison with 
$\mathcal{C}^{-1}$}, as presented in Fig. \ref{fig: ism}.
We find that in WNM {and CNM}, 
as {\it Condition (1)} is not satisfied in the entire inertial range $[L, l_\nu]$ of turbulence,
the dynamo does not go through the damping stage, but instead has a nonlinear stage (see below). 
In MC and DC, 
the damping stage of dynamo can arise at $l_\nu$,
but cannot proceed to $L$ as {\it Condition (1)} at $L$ is not met. 
Therefore, the dynamo has both damping and nonlinear stages.

The critical damping scale where the damping stage terminates can be determined by 
\begin{equation}\label{eq: trdnosc}
     \frac{2}{\mathcal{C} \Gamma_{d, \text{cr}}} = 1, 
\end{equation}
where 
\begin{equation}
    \Gamma_{d, \text{cr}} = L^{-\frac{1}{3}} V_L l_{d,\text{cr}}^{-\frac{2}{3}}.
\end{equation}
It yields (XL16)
\begin{equation}\label{eq: ldcr}
     l_{d,\text{cr}} = \Big(\frac{\mathcal{C}}{2}\Big)^\frac{3}{2} L^{-\frac{1}{2}} V_L^\frac{3}{2}.
\end{equation}
By inserting the above expression in Eq. \eqref{eq: ldevlt}, we obtain the timescale of damping stage, 
\begin{equation}
    \tau_\text{dam} =   t(l_d = l_{d,\text{cr}}) - t_1(l_{d1}=l_\nu) = \frac{23}{3} \Big(\frac{\mathcal{C}}{2} - \Gamma_\nu^{-1}   \Big) .
\end{equation}
Here we assume that {the initial magnetic field is sufficiently weak and thus} 
the damping stage starts from $l_\nu$.
The values of $l_{d,\text{cr}}$ and $\tau_\text{dam}$ for MC and DC are listed in Table \ref{Tab: ism}.

At the end of damping stage, $\mathcal{E}_M$ becomes 
\begin{equation}
        \mathcal{E}_{M,\text{dam}} = \frac{1}{2} v_{d,\text{cr}}^2 = \frac{1}{2} V_L^2 L^{-\frac{2}{3}} l_{d,\text{cr}}^\frac{2}{3}.
\end{equation}
By inserting Eq. \eqref{eq: ldcr} into the above equation, we obtain (XL16)
\begin{equation}
        \mathcal{E}_{M,\text{dam}} = \frac{\mathcal{C}}{4} L^{-1} V_L^3, 
\end{equation}
which can also be derived by combining Eq. \eqref{eq: lde} with Eq. \eqref{eq: ldcr}. 
The corresponding field strength is 
\begin{equation}
     B_\text{dam} = \sqrt{8\pi\rho_\text{eff}  \mathcal{E}_{M,\text{dam}} }.
\end{equation}
{According to Eq. \eqref{eq: trdnosc}, the ion-neutral coupling becomes strong at the end of damping stage. 
By using $\rho_\text{eff} = \rho$ in the above expression, we determine the values of 
$B_\text{dam}$, as presented in Table \ref{Tab: ism}.}
We see that 
due to the small length scale, the short timescale, and the resulting weak magnetic field, 
the damping stage is not important for the dynamo process induced by the interstellar turbulence in the partially ionized ISM.

After the short damping stage, the turbulent dynamo enters the nonlinear regime.
{Both the dynamo stretching and turbulent diffusion} of magnetic fields mainly take place at $l_p$ of $M(k,t)$, where (XL16)
\begin{equation}\label{eq: nonps}
      \Gamma_p \mathcal{E}_M  = \frac{1}{2} L^{-1} V_L^3. 
\end{equation}
By comparing $\omega_\text{IN}$ at $l_p$ with $\Gamma_p$ (Eqs. \eqref{eq: omin} and \eqref{eq: nonps}), 
\begin{equation}
     \frac{\omega_\text{IN} (l = l_p)}{\Gamma_p} = \frac{\mathcal{C}l_p^{-2}\mathcal{E}_M}{\Gamma_p}
     = \frac{\mathcal{C} \Gamma_p }{2},
\end{equation}
we see that since {\it Condition (1)} breaks down in the nonlinear stage, the above ratio is 
less than unity. 
{As the nonlinear turbulent dynamo is in a strongly coupled regime, 
the magnetic field diffusion due to the slippage between ions and neutrals} and the ion-neutral collisional damping 
are unimportant for the nonlinear stage of dynamo 
(XL16).

The nonlinear turbulent dynamo leads to a scale-by-scale equipartition between the turbulent energy and the magnetic energy. 
At the full saturation at $L$, 
all the turbulent energy carried by strongly coupled ions and neutrals can be converted to the magnetic energy. 
The saturated field strength $B_\text{non} = \sqrt{4\pi\rho}V_L $ at the end of nonlinear stage 
is presented in Table \ref{Tab: ism}, 
which provides the maximum magnitude of turbulent magnetic fields in the partially ionized ISM. 
These estimates are also consistent with the Zeeman measurements by 
\citet{Crut10}.
It implies that 
the nonlinear turbulent dynamo accounts for the turbulent magnetic fields observed in the ISM.

The timescale of nonlinear stage is (XL16)
\begin{equation}
    \tau_\text{non} = \frac{19}{3} \bigg(\frac{L}{V_L} - \Gamma_\nu^{-1}\bigg)
\end{equation}
in WNM and CNM, and 
\begin{equation}
   \tau_\text{non} = \frac{19}{3} \bigg(\frac{L}{V_L} - \frac{\mathcal{C}}{2}\bigg).
\end{equation}
in MC and DC (see Table \ref{Tab: ism}). 
It is approximately $6 \tau_\text{eddy}$,
which is longer than $\tau_\text{dam}$ by several orders of magnitude.

(b) Preshock turbulence in SNRs

When an SNR shock sweeps through the ISM, the preshock turbulence can be driven by the interaction between the 
cosmic-ray pressure gradient and interstellar density inhomogeneities
\citep{BJL09}. 
We consider the driving condition as 
(XL17)
\begin{equation}
       L=0.1\, \text{pc},   ~V_L=10^3 \,\text{km s}^{-1}.
\end{equation}
Here we use the characteristic scale of the density structure in the cold ISM
\citep{Hei03,go98} 
as $L$, and $V_L$ is of the order of the shock velocity. 
With a high dynamo stretching rate and {\it Condition (1)} satisfied in the entire inertial range $[L, l_\nu]$ of preshock turbulence
(see Fig. \ref{fig: isms}),
the preshock turbulent dynamo in all partially ionized phases remains in the damping stage. 
$l_{d,\text{cr}}$ in this case is equal to $L$. 
Accordingly, the damping stage has a timescale 
\begin{equation}
    \tau_\text{dam} =   t(l_d = L) - t_1(l_{d1}=l_\nu) = \frac{23}{3} \Big(\frac{L}{V_L} - \Gamma_\nu^{-1}   \Big) .
\end{equation}
Here we again assume that the dynamo starts at $l_\nu$ with sufficiently weak seed field. 
As $1/\Gamma_\nu$ is negligibly small compared with $L/V_L$, the values of $\tau_\text{dam}$ in different phases are approximately the same 
(see Table \ref{Tab: ism}). 
We note that $\tau_\text{dam}$ is sufficiently small compared to the precursor crossing time 
$\tau_c \sim (c/v_\text{sh}) L/V_L$, where $c$ and $v_\text{sh}$ represent light speed and shock velocity, respectively
(XL17).
So the $L$-scale magnetic field can be amplified within $\tau_c$.

$\mathcal{E}_{M,\text{dam}}$ at the end of damping stage is given by Eq. \eqref{eq: fine}. 
{As the damping stage of dynamo is in a weakly coupled regime,} 
we adopt $\rho_\text{eff} \sim \rho_i$ and 
present $B_\text{dam} \sim \sqrt{8\pi\rho_i  \mathcal{E}_{M,\text{dam}} }$ as the lower limit of $B_\text{dam}$ in Table \ref{Tab: ism}.
The dynamo-amplified magnetic field can confine energetic particles near the shock to facilitate the shock acceleration. 
For example, the maximum energy of cosmic rays that can be confined by the resulting preshock magnetic field in the case of MC is 
\begin{equation}
    E_\text{CR,max} = e B_\text{dam} L = 38.4 ~\text{PeV}.
\end{equation}
This already reaches the PeV knee of the cosmic ray spectrum and supports the Galactic origin of the cosmic rays below the knee. 
Besides, magnetic fields of the order of $100~ \mu$G near the shock front of SNRs are also inferred from observations 
(e.g., \citealt{Bam03, Bam05, Bamb05, Vin12}).

\section{ Ion-neutral coupling in MHD turbulence and in turbulent dynamo}

In a partially ionized medium, the coupling state between ions and neutrals is crucial for determining the damping of MHD turbulence 
and the efficiency of turbulent dynamo.

{\it MHD turbulence.} 
In the strong Alfv\'{e}nic turbulence 
with the magnetic energy in equipartition with the turbulent energy at $L$, 
there is a {\it critical balance} between the turbulent cascade rate, i.e., eddy-turnover rate, $ v_l / l_\perp $ 
and the Alfv\'{e}n wave frequency $\omega_A= V_{A,\text{eff}}/l_\|$ 
\citep{GS95}, 
where $l_\perp$ and $l_\|$ 
are the perpendicular and parallel components of the length scale with respect to the {\it local} magnetic field
\citep{LV99}. 
The anisotropic scaling resulting from the {\it critical balance} in the {\it local} reference system 
has been confirmed in both one-fluid 
(e.g.,  \citealt{CL02_PRL,CL03})
and two-fluid 
\citep{Burk15}
MHD simulations down to the dissipation scale of Alfv\'{e}nic turbulence.

The ion-neutral collisional damping of the turbulent cascade depends on the coupling state between ions and neutrals. 
As summarized in Table \ref{Tab: dam},
in the low wave-frequency range with $\omega_A < \nu_{ni}$,
Alfv\'{e}n waves with $V_{A,\text{eff}} = V_{A,\text{tot}}$ propagate in the strongly coupled ions and neutrals. 
\footnote{In the strong Alfv\'{e}nic turbulence, Alfv\'{e}n waves can only propagate over the distance of one wavelength due to their nonlinear interactions.}
By contrast, at high wave frequencies with $\omega_A > \nu_{in}$, 
ions and neutrals are essentially decoupled from each other, and 
Alfv\'{e}n waves with $V_{A,\text{eff}} = V_{Ai}$ can only propagate in ions. 
Within intermediate wave frequencies, neutrals are decoupled from ions, but ions are still collisionally coupled to neutrals. 
{Accordingly, Alfv\'{e}n waves propagating in the weakly coupled ions and neutrals have 
\citep{XLY14,Xuc16}
\begin{equation}
     \omega_A^2 = \frac{k_\|^2 V_{Ai}^2 [(1+\chi)\nu_{ni}^2 + k_\|^2 V_{Ai}^2]}{(1+\chi)^2 \nu_{ni}^2 + k_\|^2 V_{Ai}^2}
    = k_\|^2 V_{A,\text{eff}}^2,
\end{equation}
where $k = 1/l$, $\chi = \rho_n / \rho_i$, and 
\begin{equation}
   V_{A,\text{eff}} = \sqrt{\frac{ (1+\chi)\nu_{ni}^2 + k_\|^2 V_{Ai}^2 }{(1+\chi)^2 \nu_{ni}^2 + k_\|^2 V_{Ai}^2}} V_{Ai},      
\end{equation}
which depends on the length scale. 
MHD turbulence in the weak coupling regime} is subjected to the severest ion-neutral collisional damping.
As a result, both Alfv\'{e}n waves and Alfv\'{e}nic turbulent motions are damped 
in the weak coupling regime.

It is worth noting that 
the ambipolar diffusion scale frequently used in the literature 
(e.g., \citealt{Mous91})
\begin{equation}
     l_\text{AD} = \frac{V_{A,\text{tot}}}{\nu_{ni}}
\end{equation}
is only equivalent to the parallel neutral-ion decoupling scale for the anisotropic Alfv\'{e}nic turbulence. 
Since the energy of Alfv\'{e}nic turbulence cascades mainly along the direction perpendicular to the {\it local} magnetic field, 
we are concerned with the perpendicular neutral-ion decoupling scale, which is related to $l_\text{AD}$ via the 
{\it critical balance} mentioned above.

{\it Turbulent dynamo.}
Similarly, there also exist different ion-neutral coupling regimes for the turbulent dynamo, depending on the range of $\Gamma_l$ (see Table \ref{Tab: dam}). 
When $\Gamma_l < \nu_{ni}$, 
turbulence in the strongly coupled ions and neutrals induces the growth of magnetic energy, 
which can be expressed in terms of $V_{A,\text{eff}} = V_{A,\text{tot}}$. 
When $\Gamma_l > \nu_{in}$, 
neutrals are not involved in the dynamo process.
The dynamo only operates in ions and results in the growth of magnetic energy in terms of $V_{A,\text{eff}} = V_{Ai}$.
For an intermediate $\Gamma_l$ considered in this work, 
the dynamo takes place in the weakly coupled ions and neutrals and is affected by the strongest ion-neutral collisional damping. 
As a result, the dynamo has a damping stage.

\begin{table}[h]
\renewcommand\arraystretch{1.3}
\centering
\begin{threeparttable}
\caption[]{ Ion-neutral coupling in MHD turbulence and turbulent dynamo
}\label{Tab: dam} 
  \begin{tabular}{cccc}
      \toprule
    Coupling state                            &    Strong coupling                  &    Weak coupling                                    & Decoupling     \\
                                                              \midrule
     MHD turbulence       &     $ \omega_A < \nu_{ni}$     &   $ \nu_{ni} < \omega_A < \nu_{in} $     &  $\omega_A > \nu_{in}$   \\
             Turbulent dynamo               &     $\Gamma_l < \nu_{ni}$        &    $\nu_{ni} < \Gamma_l < \nu_{in} $       &  $\Gamma_l > \nu_{in}$   \\
 \bottomrule
    \end{tabular}
 \end{threeparttable}
\end{table}

Besides ion-neutral collisional damping, the viscosity in neutrals also leads to the damping of MHD turbulence 
in a partially ionized medium
\citep{LVC04}.
The parameter space for the dominance of neutral viscous damping and the appearance of the new regime of MHD turbulence 
in the sub-viscous range 
\citep{CLV_newregime,CLV03}
is provided in 
\citet{XLr17}.
In the context of turbulent dynamo, 
the damping stage of dynamo can only arise when the ion-neutral collisional damping 
is stronger than the neutral viscous damping.

\section{Summary}

We have studied the turbulent dynamo in a weakly ionized medium and 
numerically tested the damping stage of dynamo as theoretically predicted by XL16. 
Here we summarize our main results.

\begin{itemize}

\item We have explicitly provided the physical conditions under which the damping stage of dynamo can arise in a partially ionized medium.  
They are Eq. \eqref{eq: conlen} and Eq. \eqref{eq: con2},
and Eq. \eqref{eq: edinl} for the damping stage to persist until the damping scale reaches the injection scale $L$ of turbulence.
With sufficiently small ionization fraction and seed magnetic field, the 
timescale of damping stage is around eight times the largest eddy-turnover time
(Section \ref{sec: damthe}).

\item By performing the two-fluid dynamo simulation under the above conditions (Eqs. \eqref{eq: conlen}, \eqref{eq: con2}, and \eqref{eq: edinl})
and quantitative comparisons between the theoretical predictions and numerical measurements, 
we have numerically confirmed 
the linear-in-time growth of magnetic field strength due to the severe ion-neutral collisional damping
in the damping stage of dynamo. 
As a result of the weak coupling between ions and neutrals, most turbulent kinetic energy contained in neutrals 
cannot be converted to the magnetic energy
(Section \ref{sec: num}).

\item We have examined the physical conditions for the damping stage of dynamo in the partially ionized ISM 
and provided the parameter space for its appearance (Section \ref{sec: phsim}). 
For the dynamo induced by the interstellar turbulence, the damping stage contributes insignificantly to the dynamo growth of magnetic energy.
Instead, the nonlinear stage is mainly responsible for the growth of the interstellar turbulent magnetic fields. 
By contrast, the dynamo induced by the preshock turbulence in SNRs remains in the damping stage, 
which is important for studying the 
magnetic field amplification and cosmic ray acceleration at shocks.

\end{itemize}

S.X. acknowledges the support for Program number HST-HF2-51400.001-A provided by NASA through a grant from the Space Telescope Science Institute, which is operated by the Association of Universities for Research in Astronomy, Incorporated, under NASA contract NAS5-26555.
S.X. also thanks Chris McKee for useful conversations. 
DSB acknowledges support via NSF grants NSF-ACI-1533850, NSF-DMS-1622457, NSF-ACI-1713765 and NSF-DMS-1821242. Support from a grant by Notre Dame International is also acknowledged. 
A.L. acknowledges the support from grant
NSF-DMS-1622353.

\bibliographystyle{apj.bst}
\bibliography{xu}

\begin{thebibliography}{66}
\expandafter\ifx\csname natexlab\endcsname\relax\def\natexlab#1{#1}\fi

\bibitem[{{Balsara}(1996)}]{Bals96}
{Balsara}, D.~S. 1996, \apj, 465, 775

\bibitem[{{Balsara}(1998{\natexlab{a}})}]{Bal98a}
---. 1998{\natexlab{a}}, \apjs, 116, 119

\bibitem[{{Balsara}(1998{\natexlab{b}})}]{Bal98b}
---. 1998{\natexlab{b}}, \apjs, 116, 133

\bibitem[{{Balsara}(2004)}]{Bal04}
---. 2004, \apjs, 151, 149

\bibitem[{{Balsara}(2010)}]{Bal10}
---. 2010, Journal of Computational Physics, 229, 1970

\bibitem[{{Balsara}(2012)}]{Bal12}
---. 2012, Journal of Computational Physics, 231, 7476

\bibitem[{{Balsara} \& {Spicer}(1999{\natexlab{a}})}]{BS99a}
{Balsara}, D.~S., \& {Spicer}, D. 1999{\natexlab{a}}, Journal of Computational
  Physics, 148, 133

\bibitem[{{Balsara} \& {Spicer}(1999{\natexlab{b}})}]{BS99b}
{Balsara}, D.~S., \& {Spicer}, D.~S. 1999{\natexlab{b}}, Journal of
  Computational Physics, 149, 270

\bibitem[{{Bamba} {et~al.}(2005{\natexlab{a}}){Bamba}, {Yamazaki}, \&
  {Hiraga}}]{Bamb05}
{Bamba}, A., {Yamazaki}, R., \& {Hiraga}, J.~S. 2005{\natexlab{a}}, \apj, 632,
  294

\bibitem[{{Bamba} {et~al.}(2003){Bamba}, {Yamazaki}, {Ueno}, \&
  {Koyama}}]{Bam03}
{Bamba}, A., {Yamazaki}, R., {Ueno}, M., \& {Koyama}, K. 2003, \apj, 589, 827

\bibitem[{{Bamba} {et~al.}(2005{\natexlab{b}}){Bamba}, {Yamazaki}, {Yoshida},
  {Terasawa}, \& {Koyama}}]{Bam05}
{Bamba}, A., {Yamazaki}, R., {Yoshida}, T., {Terasawa}, T., \& {Koyama}, K.
  2005{\natexlab{b}}, \apj, 621, 793

\bibitem[{{Beresnyak}(2012)}]{Bere11}
{Beresnyak}, A. 2012, Physical Review Letters, 108, 035002

\bibitem[{{Beresnyak} {et~al.}(2009){Beresnyak}, {Jones}, \&
  {Lazarian}}]{BJL09}
{Beresnyak}, A., {Jones}, T.~W., \& {Lazarian}, A. 2009, \apj, 707, 1541

\bibitem[{{Brandenburg} \& {Subramanian}(2005)}]{Bran05}
{Brandenburg}, A., \& {Subramanian}, K. 2005, \physrep, 417, 1

\bibitem[{{Burkhart} {et~al.}(2015){Burkhart}, {Lazarian}, {Balsara}, {Meyer},
  \& {Cho}}]{Burk15}
{Burkhart}, B., {Lazarian}, A., {Balsara}, D., {Meyer}, C., \& {Cho}, J. 2015,
  \apj, 805, 118

\bibitem[{{Cho} \& {Lazarian}(2002)}]{CL02_PRL}
{Cho}, J., \& {Lazarian}, A. 2002, Physical Review Letters, 88, 245001

\bibitem[{{Cho} \& {Lazarian}(2003)}]{CL03}
---. 2003, \mnras, 345, 325

\bibitem[{{Cho} {et~al.}(2002{\natexlab{a}}){Cho}, {Lazarian}, \&
  {Vishniac}}]{CLV_newregime}
{Cho}, J., {Lazarian}, A., \& {Vishniac}, E.~T. 2002{\natexlab{a}}, \apjl, 566,
  L49

\bibitem[{{Cho} {et~al.}(2002{\natexlab{b}}){Cho}, {Lazarian}, \&
  {Vishniac}}]{CLV_incomp}
---. 2002{\natexlab{b}}, \apj, 564, 291

\bibitem[{{Cho} {et~al.}(2003){Cho}, {Lazarian}, \& {Vishniac}}]{CLV03}
---. 2003, \apj, 595, 812

\bibitem[{{Cho} \& {Vishniac}(2000)}]{Cho00}
{Cho}, J., \& {Vishniac}, E.~T. 2000, \apj, 538, 217

\bibitem[{{Cho} {et~al.}(2009){Cho}, {Vishniac}, {Beresnyak}, {Lazarian}, \&
  {Ryu}}]{CVB09}
{Cho}, J., {Vishniac}, E.~T., {Beresnyak}, A., {Lazarian}, A., \& {Ryu}, D.
  2009, \apj, 693, 1449

\bibitem[{{Crutcher} {et~al.}(2010){Crutcher}, {Wandelt}, {Heiles},
  {Falgarone}, \& {Troland}}]{Crut10}
{Crutcher}, R.~M., {Wandelt}, B., {Heiles}, C., {Falgarone}, E., \& {Troland},
  T.~H. 2010, \apj, 725, 466

\bibitem[{{Draine}(1986)}]{Dra86}
{Draine}, B.~T. 1986, \mnras, 220, 133

\bibitem[{{Draine} \& {Lazarian}(1998)}]{Dra98}
{Draine}, B.~T., \& {Lazarian}, A. 1998, \apjl, 494, L19

\bibitem[{{Draine} {et~al.}(1983){Draine}, {Roberge}, \& {Dalgarno}}]{Drai83}
{Draine}, B.~T., {Roberge}, W.~G., \& {Dalgarno}, A. 1983, \apj, 264, 485

\bibitem[{{Falkovich}(1994)}]{Fal94}
{Falkovich}, G. 1994, Physics of Fluids, 6, 1411

\bibitem[{{Goldreich} \& {Sridhar}(1995)}]{GS95}
{Goldreich}, P., \& {Sridhar}, S. 1995, \apj, 438, 763

\bibitem[{{Goodman} {et~al.}(1998){Goodman}, {Barranco}, {Wilner}, \&
  {Heyer}}]{go98}
{Goodman}, A.~A., {Barranco}, J.~A., {Wilner}, D.~J., \& {Heyer}, M.~H. 1998,
  \apj, 504, 223

\bibitem[{{Han}(2017)}]{Han17}
{Han}, J.~L. 2017, \araa, 55, 111

\bibitem[{{Haugen} {et~al.}(2004){Haugen}, {Brandenburg}, \& {Dobler}}]{Hau04}
{Haugen}, N.~E., {Brandenburg}, A., \& {Dobler}, W. 2004, \pre, 70, 016308

\bibitem[{{Haugen} {et~al.}(2003){Haugen}, {Brandenburg}, \& {Dobler}}]{Hau03}
{Haugen}, N.~E.~L., {Brandenburg}, A., \& {Dobler}, W. 2003, \apjl, 597, L141

\bibitem[{{Heiles} \& {Troland}(2003)}]{Hei03}
{Heiles}, C., \& {Troland}, T.~H. 2003, \apj, 586, 1067

\bibitem[{{Kazantsev}(1968)}]{Kaza68}
{Kazantsev}, A.~P. 1968, Soviet Journal of Experimental and Theoretical
  Physics, 26, 1031

\bibitem[{{Kowal} {et~al.}(2009){Kowal}, {Lazarian}, {Vishniac}, \&
  {Otmianowska-Mazur}}]{KL09}
{Kowal}, G., {Lazarian}, A., {Vishniac}, E.~T., \& {Otmianowska-Mazur}, K.
  2009, \apj, 700, 63

\bibitem[{{Kowal} {et~al.}(2012){Kowal}, {Lazarian}, {Vishniac}, \&
  {Otmianowska-Mazur}}]{KL12}
---. 2012, Nonlinear Processes in Geophysics, 19, 297

\bibitem[{{Kulsrud} \& {Anderson}(1992)}]{KulA92}
{Kulsrud}, R.~M., \& {Anderson}, S.~W. 1992, \apj, 396, 606

\bibitem[{{Langer}(1978)}]{Lan78}
{Langer}, W.~D. 1978, \apj, 225, 95

\bibitem[{{Lazarian} \& {Vishniac}(1999)}]{LV99}
{Lazarian}, A., \& {Vishniac}, E.~T. 1999, \apj, 517, 700

\bibitem[{{Lazarian} {et~al.}(2004){Lazarian}, {Vishniac}, \& {Cho}}]{LVC04}
{Lazarian}, A., {Vishniac}, E.~T., \& {Cho}, J. 2004, \apj, 603, 180

\bibitem[{{Li} {et~al.}(2008){Li}, {McKee}, {Klein}, \& {Fisher}}]{Li08}
{Li}, P.~S., {McKee}, C.~F., {Klein}, R.~I., \& {Fisher}, R.~T. 2008, \apj,
  684, 380

\bibitem[{{Lithwick} \& {Goldreich}(2001)}]{LG01}
{Lithwick}, Y., \& {Goldreich}, P. 2001, \apj, 562, 279

\bibitem[{{Maron} \& {Blackman}(2002)}]{Mar02}
{Maron}, J., \& {Blackman}, E.~G. 2002, \apjl, 566, L41

\bibitem[{{Maron} {et~al.}(2004){Maron}, {Cowley}, \& {McWilliams}}]{Ma04}
{Maron}, J., {Cowley}, S., \& {McWilliams}, J. 2004, \apj, 603, 569

\bibitem[{{Maron} \& {Goldreich}(2001)}]{MG01}
{Maron}, J., \& {Goldreich}, P. 2001, \apj, 554, 1175

\bibitem[{{Meyer} {et~al.}(2014){Meyer}, {Balsara}, {Burkhart}, \&
  {Lazarian}}]{Mey14}
{Meyer}, C.~D., {Balsara}, D.~S., {Burkhart}, B., \& {Lazarian}, A. 2014,
  \mnras, 439, 2197

\bibitem[{{Mouschovias}(1991)}]{Mous91}
{Mouschovias}, T.~C. 1991, \apj, 373, 169

\bibitem[{{Oishi} \& {Mac Low}(2006)}]{Ois06}
{Oishi}, J.~S., \& {Mac Low}, M.-M. 2006, \apj, 638, 281

\bibitem[{{O'Sullivan} \& {Downes}(2006)}]{Os06}
{O'Sullivan}, S., \& {Downes}, T.~P. 2006, \mnras, 366, 1329

\bibitem[{{O'Sullivan} \& {Downes}(2007)}]{Os07}
---. 2007, \mnras, 376, 1648

\bibitem[{{Schekochihin} {et~al.}(2002){Schekochihin}, {Maron}, {Cowley}, \&
  {McWilliams}}]{Sc02}
{Schekochihin}, A.~A., {Maron}, J.~L., {Cowley}, S.~C., \& {McWilliams}, J.~C.
  2002, \apj, 576, 806

\bibitem[{{Shu}(1992)}]{Shu92}
{Shu}, F.~H. 1992, {The physics of astrophysics. Volume II: Gas dynamics.}

\bibitem[{{Spitzer}(1978)}]{Spi78}
{Spitzer}, L. 1978, {Physical processes in the interstellar medium}

\bibitem[{{Subramanian}(1998)}]{Sub98}
{Subramanian}, K. 1998, \mnras, 294, 718

\bibitem[{{Tilley} \& {Balsara}(2008)}]{TB08}
{Tilley}, D.~A., \& {Balsara}, D.~S. 2008, \mnras, 389, 1058

\bibitem[{{Tilley} \& {Balsara}(2010)}]{TilBal10}
---. 2010, \mnras, 406, 1201

\bibitem[{{Tilley} \& {Balsara}(2011)}]{Till11}
---. 2011, \mnras, 415, 3681

\bibitem[{{Tilley} {et~al.}(2012){Tilley}, {Balsara}, \& {Meyer}}]{Til12}
{Tilley}, D.~A., {Balsara}, D.~S., \& {Meyer}, C. 2012, New Astronomy, 17, 368

\bibitem[{{Vink}(2012)}]{Vin12}
{Vink}, J. 2012, \aapr, 20, 49

\bibitem[{{Vranjes} \& {Krstic}(2013)}]{VrKr13}
{Vranjes}, J., \& {Krstic}, P.~S. 2013, \aap, 554, A22

\bibitem[{{Xu} \& {Lazarian}(2016)}]{XL16}
{Xu}, S., \& {Lazarian}, A. 2016, \apj, 833, 215

\bibitem[{{Xu} \& {Lazarian}(2017{\natexlab{a}})}]{XL17}
---. 2017{\natexlab{a}}, \apj, 850, 126

\bibitem[{{Xu} \& {Lazarian}(2017{\natexlab{b}})}]{XLr17}
---. 2017{\natexlab{b}}, New Journal of Physics, 19, 065005

\bibitem[{{Xu} {et~al.}(2015){Xu}, {Lazarian}, \& {Yan}}]{XLY14}
{Xu}, S., {Lazarian}, A., \& {Yan}, H. 2015, \apj, 810, 44

\bibitem[{{Xu} {et~al.}(2016){Xu}, {Yan}, \& {Lazarian}}]{Xuc16}
{Xu}, S., {Yan}, H., \& {Lazarian}, A. 2016, \apj, 826, 166

\bibitem[{{Zaqarashvili} {et~al.}(2011){Zaqarashvili}, {Khodachenko}, \&
  {Rucker}}]{Zaqa11}
{Zaqarashvili}, T.~V., {Khodachenko}, M.~L., \& {Rucker}, H.~O. 2011, \aap,
  529, A82

\end{thebibliography}

\end{document}